\documentclass[aps,prl,amssymb]{revtex4}
\usepackage{graphicx}
\usepackage{dcolumn}
\usepackage[T2A]{fontenc}
\usepackage[cp1251]{inputenc}

\begin{document}
\title{LOCALIZED --- DELOCALIZED ELECTRON QUANTUM PHASE TRANSITIONS}
\author{V.F.Gantmakher, V.T.Dolgopolov}

\affiliation{Institute of Solid State Physics RAS, Chernogolovka
142432, Russia\\
 tel. +7-496-5225425, +7-496-5222946,\\ fax +7-496-5249701\\
 E-mail: gantm@issp.ac.ru, dolgop@issp.ac.ru}

\begin{abstract}

Metal--insulator transitions and transitions between different quantum
Hall liquids are used to describe the physical ideas forming the basis of
quantum phase transitions and the methods of application of theoretical
results in processing experimental data.  The following two theoretical
schemes are discussed and compared: the general theory of quantum phase
transitions, which has been developed according to the theory of
thermodynamic phase transitions and relies on the concept of a partition
function, and  a theory which is based on a scaling hypothesis and the
renormalization-group concept borrowed from quantum electrodynamics, with
the results formulated in terms of flow diagrams.
\end{abstract}
 \maketitle

\noindent
 CONTENT

\vspace{2mm}
 \noindent{\bf 1. Introduction}

1.1.  Thermal and quantum fluctuations.\\
{\bf 2. Quantum phase transitions}

2.1. Definition of a quantum phase transition . 2.2. Parallels and
differences between classical and quantum phase transitions 2.3. Critical
region of a quantum phase transition.\\
{\bf 3. Flow diagrams for metal--insulator transitions}\\
{\bf 4. Three-dimensional electron gas}\\
{\bf 5. Two-dimensional electron gas }

5.1. Gas of noninteracting electrons  5.2. Spin--orbit interaction 5.3.
Gas of interacting electrons.\\
{\bf 6. Quantum transitions between the different states of a Hall liquid}\\
{\bf 7. Conclusion}

\section{1. Introduction}

Quantum phase transitions that occur in an electron gas and result in
localization represent the mainstream in the study of electrons in solids; it
is directed toward low temperatures and interactions. We think that a dangerous
gap between theory and experiment has appeared in this field of science. The
related theories are so complex that experimentalists, as rule, use ready
recipes and formulas of these theories without knowing (and, hence, without
controlling) all the initial assumptions and limitations.

In this review, we will try to close this gap. Our review does not contain a
sequential description of the mathematical techniques involved in formulating
hypotheses and theories. Instead, we outline the physical ideas, concepts, and
assumptions that are often omitted in theoretical works and reviews, especially
at the stage of a developed theory. This specific feature is thought to make
our review interesting and useful for experimentalists.

On the other hand, we did not tend to accumulate, classify, and estimate the
huge experimental data obtained in this field. We address experiment to
demonstrate the use of a theory for its interpretation and the related
difficulties and problems. This specific feature is thought to make our review
interesting and useful for theorists.

We first qualitatively discuss the ides that constitute the basis of the theory
of quantum phase transitions. We then consider how the general theoretical
scheme can describe the well-known data on metal--insulator transitions and
transitions between different quantum Hall liquids.

\vspace{5mm}{\bf 1.1.  Thermal and quantum fluctuations}

Before discussing quantum phase transitions in essence, we briefly recall some
facts from statistical physics. We consider a macroscopic system consisting of
a huge number of particles. This system is almost isolated and behaves mainly
as a closed system; however, it can exchange particles and energy with a larger
system, which serves as a reservoir. In other words, our system is a subsystem
of this reservoir.

We first analyze the classical subsystem at a finite temperature T. According
to classical statistics formulas, the probability $p_i$ for the subsystem to
occupy the state with energy $\varepsilon_i$ is
\begin{equation}\label{a}
  p_i=\frac{e^{-\varepsilon_i/T}}{Z}\:,
\end{equation}
where the coefficient $1/Z$ is determined by the normalization condition
 $\sum p_i=1$
\begin{equation}\label{b}
Z=\sum_i\exp(-\varepsilon_i/T)
\end{equation}
(hereafter, temperature is given in energy units). Function $Z$ is called
the partition function. It plays a key role in the description of the
thermodynamic properties of classical objects. That the subsystem energy
obeys distribution (\ref{a}) rather than being fixed means the presence of
classical thermal fluctuations.

When quantum mechanics appeared, the basis of classical thermodynamics and its
principal formulas were revised. The classical expressions were found to have a
limited field of application. Quantum statistics requires a developed
quantum-mechanics math-based environment. Before addressing this environment,
we assume that the subsystem is completely closed and does not interact with
the large system. As a result, we can use the standard quantum mechanics
technique and use the Schr\"{o}dinger steady-state equation to describe the
subsystem. The resulting set of steady-state energies $\varepsilon _i$ and the
corresponding wavefunctions $\varphi _i(q)$ of the subsystem are considered to
be subsystem attributes in the zeroth approximation. These energies,
$\varepsilon _i$, enter Eqn (\ref{b}). The set of $\varphi _i(q)$ functions is
convenient because it is related to the subsystem under study and is a complete
set; therefore, it can be used for expansion into a series.

The presence of a huge number of particles in the subsystem implies that it has
a very dense energy distribution of quantum levels. Due to a weak interaction
with the reservoir, the subsystem is in the so-called mixed state, in which no
measurement or a set of measurements can lead to unambiguously predicted
results. Upon determining the mixed state, the repeated measurements of any
physical quantity give a near-average result that differs from the previous
one. The scatter of the measured physical quantities is interpreted as a result
of fluctuations.

Because the subsystem interacts with the environment, it cannot be described by
a wave function; therefore, the wave function is substituted by a density
matrix,
\begin{equation}\label{aa}
\rho(q',q) = \int\psi^*(q',X )\psi(q,X) dX,
\end{equation}
where $q$ corresponds to the set of subsystem coordinates, $X$ are the
remaining coordinates of the system, and $\psi (q, X\,)$ is the wave function
of the closed system.

The mean values of any physical quantity $\langle s\rangle$ are now calculated
using the density matrix
\begin{equation}\label{ab}
 \langle s\rangle =\frac1A \int\left\{\widehat s[\rho(q',q)]
 \rule[-2mm]{0pt}{12pt}\right\}
 _{q'=q}dq,\qquad
 A=\int\rho(q,q)dq.
\end{equation}
rather than a wave function.  In Eqn (\ref{ab}), we apply the operator
$\widehat s$, which acts on functions of the variable $q$, to the density
matrix $\rho (q\,',q)$; we then set $q\,'=q$ and integrate. The operator that
formally enters the expression for the normalization factor $A$ is identical
with unity; therefore, the condition $q\,'=q$ is naturally taken into account
in the integrand.

According to general rule(\ref{ab}), the mean of coordinate $\langle q\rangle$,
for example, is given by
\begin{equation}\label{q}
 \langle q\rangle =\frac1A\int\int\psi^*(q,X)q\psi(q,X)dqdX=
 \frac{\int q\rho(q,q)dq}{\int\rho(q,q)dq}.
\end{equation}

The physical meaning of the density matrix can be clarified if we write it in
an explicit form for the subsystem in statistical equilibrium at a finite
temperature $T$:
\begin{equation}\label{Dm}
 \rho(q',q) = \sum_i\varphi_i^*(q')\varphi_i(q)e^{-\varepsilon_i/T}.
\end{equation}
For the subsystem in statistical equilibrium, it follows from Eqns (\ref{q})
and (\ref{Dm})
\begin{equation}\label{q+Dm}                    
 \langle q\rangle=
 \frac1A\sum_i\int\varphi_i^*(q)\varphi_i(q)e^{-\varepsilon_i/T}qdq
 =\frac1A\sum_i\langle q\rangle_ie^{-\varepsilon_i/T},\qquad
 {\langle q\rangle}_i=\int\varphi_i^*(q)\varphi_i(q)qdq.
\end{equation}
The averaging with the density matrix performed in Eqn (\ref{q+Dm}) account for
both the probabilistic description in the form of $\langle q\rangle _i$ in
quantum mechanics and incomplete information about the system (statistical
averaging).

We use the $\varphi _i(q)$ functions and write the function $\rho (q\,',q)$ in
the matrix form
\begin{equation}\label{Mrho}                        
 \rho(q',q)=\|\rho_{ij}\|=
 \left\|\int\varphi_j^*(q)\rho(q',q)\varphi_i(q)dq\right\|.
\end{equation}
In Eqn (\ref{ab}), we then have
\begin{equation}\label{ad}
 \langle s\rangle=\frac{\sum_{ij}s_{ij}\rho_{ji}}{\sum_i\rho_{ii}}
 =\sum_{ij}s_{ij}w_{ji},\qquad
 w_{ij}=\frac{\rho_{ij}}{\sum_i\rho_{ii}}.
\end{equation}
The $w_{i\,j}$ matrix constructed from the set of $\varphi _i(q)$ functions is
called the statistical matrix and, in essence, is the normalized density
matrix.

We replace $\widehat s$ with the energy operator $\widehat H$ in Eqn (\ref{ab})
and obtain
\begin{equation}\label{e}
 \langle\varepsilon\rangle =\sum_iw_{ii}\varepsilon_i.
\end{equation}
This means that the probability $p_i$ of detecting the energy $\varepsilon _i$
in the subsystem is equal to the diagonal element of the matrix $w_{ii}$,
\begin{equation}
\label{c}
 p_i = {w}_{ii}.
\end{equation}
Equation (\ref{c}) is the quantum analog of Eqn (\ref{a}).

The statistical matrix has a number of universal properties. By definition,
this matrix is normalized:
\begin{equation}\label{w}
\sum_i w_{ii} = 1.
\end{equation}
As follows from Eqn (\ref{Dm}), the statistical matrix is diagonal in
statistical equilibrium. Its diagonal elements are functions of only the energy
of the corresponding subsystem state $\varepsilon _i$. Until at least one
$w_{ii}$ with $i\neq 0$ has a nonzero value, the subsystem state remains mixed,
and any measured quantity undergoes fluctuations.

The macroscopic system is in a mixed state mainly due to a finite temperature.
A comparison of Eqns (\ref{a}) and (\ref{c}) demonstrates that in the
high-temperature limit, the statistical matrix obeys the Gibbs distribution

\begin{equation}\label{1}
  w_{ii}=\exp(-\varepsilon_i/T)/Z.
\end{equation}
with a good accuracy.

Because the integral of the squared wave function $\varphi _i(q)$ over the
entire coordinate space is
 $$\int\varphi_i^*(q)\varphi_i(q)dq\equiv1,$$
the partition function $Z$ can be represented as the sum of the diagonal matrix
elements of the operator $\exp {(-\widehat {H}/T\,)}$ over the complete set of
eigenfunctions $\varphi _i$:

\begin{equation}\label{qpt3}
 Z=
 \sum\limits_i\int\varphi_i^*(q)\varphi_i(q)e^{-\varepsilon_i/T}dq
 =\sum_i\langle\varphi_i|\exp(-\widehat{H}/T)|\varphi_i\rangle.
\end{equation}
Here, we use that $\varphi _i$ are eigenfunctions of the Hamiltonian $\widehat
H$ to obtain $$\exp(-\widehat{H}/T)\varphi_i=e^{-\varepsilon_i/T}\varphi_i$$.

A comparison of Eqns (\ref{ab})---(\ref{q+Dm}) with Eqn (\ref{1}) gives another
form of the partition function,

\begin{equation}\label{bb}                  
Z=\int\rho(q,q)dq\equiv\mbox{Sp}[\rho(q',q)].
\end{equation}
If $Z$ is regarded as a normalization factor, Eqns (\ref{ab}) and (\ref{ad})
demonstrate that representation (\ref{bb}) is valid in both the classical limit
and general case.

\begin{figure}[h]           
\includegraphics{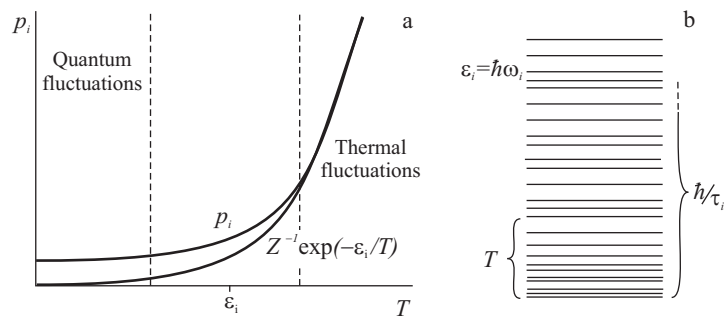}
\caption{(a) Low-temperature behavior of the diagonal elements $w_{ii}$ of
the statistical matrix corresponding to low energies. (b) Simultaneous
excitation of thermal and quantum fluctuations } \label{Spector1}
\end{figure}

As the subsystem is open, it interacts with the environment. Even if this
interaction is very weak, the states of the macroscopic system become mixed in
a certain small energy range due to the extremely high density of the system
energy levels. Therefore, even at zero temperature, the unclosed subsystem is
in a mixed state. This means that as the temperature decreases, each $w_{ii}\,$
$(i\ne 0)$ element approaches a temperature-independent constant, which depends
on $\varepsilon _i$ and the sizes and properties of the subsystem, rather than
tending to zero exponentially. Thermal fluctuations are replaced by quantum
fluctuations. This statement is illustrated by Fig.\,\ref{Spector1}a, which
qualitatively shows the temperature dependence of the probability $p_i$ of
detecting the energy $\varepsilon _i$ in the subsystem. The dashed lines
separate the regions of predominantly quantum and predominantly thermal
fluctuations. The limiting value $p_i(0)$ is specified by the energy
$\varepsilon _i$, sizes, and properties of the subsystem.

In this subsystem, $p_i(0)$ obviously decreases with an increase in
$\varepsilon _i$. We assume that this dependence is a power-law function,
$p_i(0)\propto 1/\varepsilon _i^{\,\alpha }$, and obtain the lower estimate of
the exponent $\alpha$. Energy for thermal fluctuations is taken from the
reservoir. A finite, not exponentially small, probability of detecting the
subsystem in a state $\varepsilon _i\gg T$ means a violation of the energy
conservation law. This violation can exist in the framework determined by the
uncertainty relation

\begin{equation}\label{uncert}          
  \varepsilon_i\tau_i\sim\hbar,
\end{equation}
where $\tau _i$ is the lifetime of the state $\varepsilon _i$. Therefore, even
if transitions to all $\varepsilon _i$ states were equiprobable, $p_i(0)$ would
decrease according to the law $p_i(0)\propto \varepsilon _i^{\,-1}$, as follows
from Eqn (\ref{uncert}) . As a result, we have
\begin{equation}\label{p_i}             
 \alpha\geqslant1.
\end{equation}
It is essential that the energy range in which quantum fluctuations occur is
temperature independent, which follows from uncertainty relation
(\ref{uncert}).

The relative role of thermal and quantum fluctuations is illustrated in
Fig.\,\ref{Spector1}b, which shows the excitation spectrum of the
subsystem in one of the phases, i.e., far from phase transition point
$x=x_{\rm c}$. The brace in the bottom left part of Fig.\,\ref{Spector1}b
indicates the temperature-related scale. Only modes with frequencies
$\omega _i\,\lesssim \,T/\hbar$  are classically excited modes. Although
the range of thermal modes is bounded above by temperature, they have
large occupation numbers. Modes with $\omega _i>T/\hbar$ are mainly
excited due to quantum processes and have small occupation numbers.
However, as the temperature decreases, the role of quantum excitations
increases.
\newpage

\section{2. Quantum phase transitions}

{\bf 2.1 Definition of a quantum phase transition.}

Quantum phase transitions are phase transitions that can occur at the absolute
zero $T=0$ and consist of a change in \emph{the ground state} of a system in
the case where a certain control parameter $x$ takes a critical value $x_{\rm
c}$. (The ground state is taken to be the lowest possible mixed state.) The
control parameter can be, for instance, a magnetic field or an electron
concentration. Quantum phase transitions belong to the class of continuous
transitions at which none of the physical functions of state has a
discontinuity at the transition point.

A quantum fluctuation is the only cause that can change the ground state of the
system at zero temperature; the phase transitions under study are therefore
called quantum phase transitions. As in the case of thermodynamic transitions,
the concept of a correlation length $\xi$, which has the meaning of the average
quantum-fluctuation size, is introduced into the theory of quantum phase
transitions. At the absolute zero temperature, $\xi$ is only determined by the
deviation $\delta x$ of the control parameter from the critical value. The
dependence of $\xi$  on $\delta x$ is assumed to be a power-law function,

\begin{equation}\label{dop1}   
   \xi\propto|x-x_c|^{-\nu}.
\end{equation}
Real experiments are always performed at finite temperatures, where thermal
fluctuations exist in addition to quantum fluctuations. The goal of the theory
is to predict the manifestation of the phase transition that occurs only at
zero temperature in the properties of the system at a finite temperature.

The theory of quantum phase transitions (see, e.g., book \cite{Sachdev} or
reviews  \cite{QuantRev,Vojta}) is analogous to the theory of thermodynamic
phase transitions. In the $(x,T))$ plane (i.e., in a plane where temperature is
plotted versus a control parameter), the quantum transition point can be a
finite point on the line of thermodynamic transitions, $x_{\rm
c}(T\,)\rightarrow x_{\rm c}(0)$ as $T\rightarrow 0$. For example, this
behavior is characteristic of magnetic transitions with the magnetic field used
as a control parameter. Quantum phase transitions of another type also exist;
they are imaged by isolated point $x_{\rm c}$ on the abscissa axis of the
$(x,T)$ plane. The metal--insulator transition is an example of such a
transition. In this review, we restrict ourselves to the case of isolated-point
transitions.

\vspace{5mm}

{\bf 2.2 Parallels and differences between classical and quantum phase
transitions}

As noted above, quantum phase transitions belong to the class of continuous
transitions; there are no two coexisting competitive phases and, hence, no
stationary boundary between them. This class also includes thermodynamic
second-order phase transitions. At a thermodynamic phase transition point, the
system transforms into another phase as a whole as a result of thermal
fluctuations. Fluctuations exist on either side of the transition, and their
characteristic size $\xi$  is called the correlation length. As the transition
is approached from either side, $\xi$  diverges \cite{PP,Goldenfeld}.

It is natural to expect a similar situation to occur in the vicinity of a
quantum transition with the participation of quantum fluctuations. An analogy
between classical and quantum phase transitions does exist, and it is rather
unexpected. \emph{The behavior of a quantum system in the vicinity of a
transition point at a finite temperature in a d-dimensional space is analogous
to the behavior of a classical system in a space with dimension ${\cal D}>d$.}
This statement requires extensive explanations, which should include an
algorithm for the introduction of such an imaginary system and the
determination of its dimension.

First, we have to clearly distinguish between the dimension of the geometric
space in which the system is located and the dimension of the
generalized-coordinate space, which depends on the number of particles $\cal N$
in the system. If $\varrho$ is the density of particles in the space, we have
${\cal N}=\int \int \int \varrho d\,^dX$, where the dimension $d$ determines
the multiplicity of the integral. Usually, a $d$-dimensional space is assumed
to be infinite in all directions, but the range of one or several $X_i$
coordinates can be limited.

Second, we note a similarity between the operator $\exp{(-\widehat {H}/T\,)}$,
which was used to write Eqn (\ref{qpt3}), and the operator $\widehat {S}$ that
describes the evolution of a closed quantum system with time according to the
Schr\"{o}dinger equation
\begin{equation}\label{qpt4}
 {\rm i}\hbar \frac{\partial\psi}{\partial t}=\widehat{H}\psi,\qquad
 \widehat{S}=\exp(-\frac{{\rm i}}{\hbar}\widehat{H}t).
\end{equation}
This similarity becomes obvious if we change the variables as
\begin{equation}\label{qpt11}
  {\rm i}t/\hbar=1/T.                   
\end{equation}

With substitution (\ref{qpt11}), we can interpret the matrix elements in Eqn
(\ref{qpt3}) differently. The element $\bigl \langle i\,\bigl |\,\exp
{(-\widehat {H}/T\,)}\bigr |i\,\bigr \rangle$  can be regarded as the amplitude
of the probability that, starting from state $\langle i\,|$, the subsystem
evolves under the action of $\widehat {S}$ and returns to the initial state
$|i\,\rangle$  within imaginary time $\tilde t$, which is equal to $-{\rm
i}\hbar /T$. The imaginary time $\tilde t$ is often called the Matsubara time.
For definiteness, we assume that the number of steps is fixed and equal to
$N+1$, $N\gg 1$ and that, in time ${\rm i}\hbar /T$, the subsystem has passed
through $N$ virtual states and occupied each state for a time $(\delta \tilde
t\,)_j$; as a result, we have

\begin{equation}\label{qpt12}                       
\sum_{j=0}^N(\delta \tilde t\,)_j={\rm i}\hbar/T.
\end{equation}

The amplitude of `the probability of returning to the initial state' means the
sum of the amplitudes of the probabilities of returning for all possible
trajectories in the space of states. We consider a set of trajectories
consisting of $N$ steps. The operator entering each matrix element in Eqn
(\ref{qpt3}) is represented as

\begin{equation} \label{exp}        
\exp(-\widehat{H}/kT)= \exp\left(\frac{{\rm
i}}{\hbar}\widehat{H}\sum_0^N(\delta \tilde t\,)_j\right).
\end{equation}
We take a complete set of wavefunctions $|m_j\rangle$ of an arbitrary operator
that does not commute with the Hamiltonian and assume that the trajectories are
realized in these states. Then, instead of Eqn (\ref{qpt3}), we obtain

\begin{equation} \label{Z1}
Z=
 \sum\limits_i\sum\limits_{m_1,m_2,...m_N}
 <i|\exp(-{\rm i}\widehat{H}(\delta \tilde t\,)_1/\hbar)|m_1>
 <m_1|\exp(-{\rm i}\widehat{H}(\delta \tilde t\,)_2/\hbar)|m_2>...
 <m_N|\exp(-{\rm i}\widehat{H}(\delta \tilde t\,)_N/\hbar)|i>.
\end{equation}
The product of the matrix elements in the summand corresponds to a chain of
consecutive virtual transitions. The summation over all $\{m_j\}$ combinations
means that all possible closed chains of $N$ links are taken into account. The
modification of the expression for $Z$ --- the replacement of Eqn (\ref{qpt3})
with Eqn (\ref{Z1}) --- means that we take the quantum properties of the system
into account by adding virtual transitions to real ones.

\begin{figure}[h]
\includegraphics{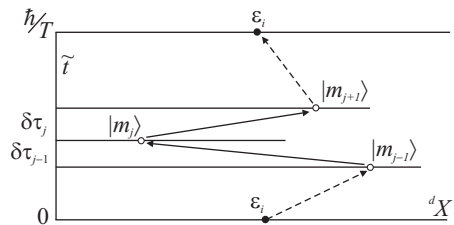}
 \caption{Set of quantum statistical $d$-dimensional systems located in a
one-dimensional band of width $\hbar/T$and virtual transition between
them} \label{d+1}
\end{figure}

The introduction of virtual transitions is illustrated in Fig. \ref{d+1}, where
the abscissa axis stands for the initial $d$-dimensional space $\{^dX\}$ and
all possible classical states of the system are located along this axis in the
order of increasing energy. All other horizontal lines are replicas of this
space. These replicas form a set of $N$ elements in the segment $[0,\hbar
/T\,]$. The black dot on the abscissa axis represents the initial state of the
system with energy $\varepsilon _i$. The second black dot on the upper
horizontal line corresponds to the `final' state of the system with the same
energy $\varepsilon _i$ at the maximum value of the imaginary time $|\tilde
{t}\,|=\hbar /T$. The arrows indicate the chain of virtual transitions through
the set of virtual states $|m_j\rangle$  designated by white dots. These are
mixed states without a definite energy; therefore, they are not eigenstates of
the operator $\widehat {H}$. The virtual transitions reflecting the quantum
properties of the statistical system under study are shown by arrows. The chain
of virtual transitions shown in Fig. \ref{d+1} corresponds to one term in the
sum in (\ref{Z1}). The summation over all $\{m_j\}$ combinations means that we
account for all possible chains between the fixed initial and final
$\varepsilon _i$ points.

Now, the problem is to construct a classical system whose partition function is
represented by Eqn (\ref{Z1}). In the original Eqn (\ref{qpt3}), the summation
over $i$ meant the summation over all the states realized in a $d$-dimensional
system. The number of terms in the sum is now increased, and our real system
does not have such a large number of different states. Nevertheless, we can
construct an imaginary classical system with a higher dimension using the
scheme shown in Fig. \ref{d+1}.

We add an additional axis for an imaginary time to the $d$ axes of the original
space. In the graphical terms of Fig. \ref{d+1}, this means that an ordinate
axis is added to the abscissa axis. It is seen from Eqn (\ref{qpt12}) that the
coordinate along this new axis changes in the range

\begin{equation}\label{qpt15}                   
0\leqslant|\widetilde{t}|\leqslant\hbar/kT.
\end{equation}
Each point in the strip (\ref{qpt15}) in Fig. \ref{d+1} corresponds to
some virtual state of the quantum system; it can be named the image point.
The width (\ref{qpt15}) of strip in which an image point is located
increases with decreasing the temperature. At $T=0$, the strip transforms
into a half-plane; an increase in the temperature, in contrast, narrows
the band and decreases its contribution to the statistical properties of
the quantum system.

We repeat the $d$-dimensional classical system $N$ times by placing replicas
along the imaginary time $\tilde t$ axis at a distance $(\delta \tilde {t})_j$
from each other. The states of the replicas in this ensemble can be different.
Let these states be $m_j$. We fix a certain `initial' state of the `lower'
subsystem $\varepsilon _i$. The energy of the classical ${\cal D}$-dimensional
system, we are constructing, is equal to the sum of the energies of all layers,
with allowance for the interaction between them. Each $\varepsilon _{i,\{m\}}$
corresponds to a certain energy $\varepsilon _i$ of the initial $d$-dimensional
classical system and a certain chain of states entering Eqn (\ref{Z1}).
Therefore, the number of terms in the sum $\sum _{\{m\}}s_{i,\,m}$ in Eqn
(\ref{Z1}) is equal to the number of different $\varepsilon _{i,\{m\}}$
energies. We note that each $s_{i,\,m}$ term in sum has a form typical for
partition sums, since the product of the matrix elements in this term
eventually reduces to the product of $\exp {\bigl (-{\rm i}\varepsilon
_j(\delta \tilde {t}\,)_j/\hbar \bigr )}$. The only difference is that for the
classical system, each term in $Z$ is a real positive number and, in Eqn
(\ref{Z1}), a real positive number is represented by the sum $s_i=\sum
_{\{m\}}s_{i,\,m}$ over $\{m\}$ of all $s_{i,\,m}$ products. When neglecting
this difference, we can consider Eqn (\ref{Z1}) the partition function $Z$ of
the imaginary ${\cal D}$-dimensional classical system.

Of course, in the general form, the resulting mathematical construction is
absolutely unpractical. However, as is customary in deriving scaling
relations, it is important to reveal some significant properties of sum
(\ref{Z1}) rather than to calculate it. As is shown in Section 2.3, one of
these properties is anisotropy, i.e.,  non equivalence of the axes of the
${\cal D}$-dimensional space.

\begin{figure}[h]                   
\includegraphics{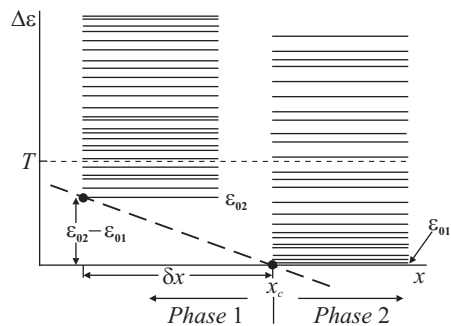}
 \caption{The difference in the lowest energies of two phases
$\Delta\varepsilon=\varepsilon_{02}-\varepsilon_{01}$ versus a control
parameter (dashed line). At a control parameter $x=x_c-\delta x$, phase 1
is subjected to fluctuation excitation inside equilibrium phase 2. Because
$|\Delta\varepsilon|<T$, thermal fluctuation play a key role in this
excitation} \label{Spector2}
\end{figure}
At a thermodynamic phase transition point, the partition function $Z$ has
special features. The sensitivity of function (\ref{qpt3}) to the presence
of a phase transition is caused by the fact that, when the transition is
approached, the energy range $T$ determining significant terms in sum
(\ref{qpt3}) contains not only $\varphi _i$ levels from the set
corresponding to the equilibrium phase but also $\varphi _i'$ levels from
the nonequilibrium phase (Fig. \ref{Spector2}). This allows fluctuation
transitions between $|i\,\rangle$  and $|i\,'\rangle$  levels from
different sets. Additional possibilities appear after Eqn (\ref{qpt3}) is
replaced with Eqn(\ref{Z1}). If the temperature is low $(|\Delta
\varepsilon |>T\,)$, the fluctuation-induced appearance of another phase
is also possible, but due to quantum fluctuations.

\vspace{5mm}{\bf 2.3. Critical region of a quantum phase transition}

A region in which all physical quantities depend only on the correlation
length $\xi$ always exists in the phase plane near a classical phase
transition. It is called the critical or scaling region. Near a quantum
phase transition at $T=0$, there also exists a control parameter $\delta
x$ range in which physical quantities are expressed through the length
$\xi$ determined by Eqn (\ref{dop1}). In Section 4, we show this behavior
by the example of a metal--insulator transition in a three-dimensional
system of noninteracting electrons, and determine the boundaries of this
region. At a finite temperature $T\neq0$, however, the scaling region of a
quantum phase transition is more complex.

The space $\{^dX,\tilde {t}\,\}$ has dimension $d+1$ because of the
additional imaginary time axis. We introduce correlation lengths in the
space $\{^dX,\tilde {t}\,\}$. We retain the traditional notation $\xi$ for
the correlation length in an ordinary $d$-dimensional subspace and let
$\xi_\varphi$ denote the correlation length along the additional axis. The
subscript $\varphi$ is a reminder that $\xi _\varphi$  is related to the
quantum aspect of the problem and to the specific features of wave
functions. The dimension of $\xi _\varphi$  is $\hbar /T$ rather than
length; that is, it is measured in seconds rather than centimeters.

As $x\rightarrow x_{\rm c}$ and $T\rightarrow 0$, both correlation lengths
diverge at a transition. According to the theory of continuous
thermodynamic transitions, the divergence is described by power functions;
but the exponents of the two correlation lengths can be different in
general. This is usually written as
\begin{equation}\label{qpt16}
   \xi\propto\delta x^{-\nu},\qquad\delta x=|x-x_c|,
\end{equation}
\begin{equation}\label{qpt17}
   \xi_\varphi\propto\xi^z.
\end{equation}
The exponents $\nu$  and $z$ are called critical indices; $z$ is called
the dynamic critical index. Both names and designations originate from the
theory of thermodynamic processes. In particular, the dynamic critical
index in the theory of thermodynamic processes enters the relation between
the lifetime and size of thermal fluctuations, $\tau \propto \xi ^{\,z}$.
In the quantum problem at $T=0$, the situation is similar: the length
$\xi$ characterizes spatial correlations, i.e., the characteristic size of
quantum fluctuations, and the time $\xi _\varphi$  characterizes time
correlations.

Formula (\ref{qpt17}) immediately fixes the dimension of the space in
which the imaginary classical system should be placed. A real length must
be the coordinate along all axes of this space; that is, length axes
should be made from the $\tilde t$ axis. As follows from dimensional
considerations and Eqn (\ref{qpt17}), the length equivalent to the
`correlation pseudolength' $\xi _\varphi$ is proportional to $\xi _\varphi
^{1/z}$,
\begin{equation}\label{qpt17a}
   \zeta\propto\xi_\varphi^{1/z}.
\end{equation}
Therefore, the volume element in the space is
    \begin{equation}\label{volume}                          
    (d\xi)^dd\xi_\varphi\propto(d\xi)^d(d\zeta)^{z}
\end{equation}
and a $z$-dimensional subspace with the conventional `spatial' coordinates
appears instead of the one-dimensional imaginary time axis. Hence, we have
\begin{equation}\label{calD}                   
   {\cal D}=d+z.
\end{equation}

\begin{figure}[h]                               
\includegraphics{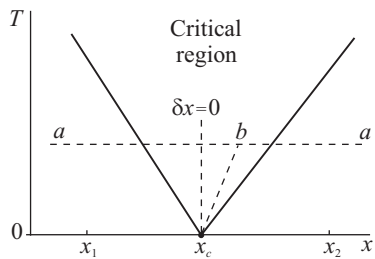}
\caption{Shape of the critical region of a quantum phase transition
depicted by an isolated point in the ($x,T$) plane} \label{Qpt1}
\end{figure}
It is convenient to discuss the consequences of Eqns (\ref{qpt16}) and
(\ref{qpt17}) using the $(x,T\,)$ plane (Fig. \ref{Qpt1}). At $T=0$, the
control parameter $x$ on the abscissa axis of the $(x,T\,)$ plane only
affects one independent correlation length, $\xi$, and $\xi _\varphi$ (and
the length $\zeta$) can be formally obtained from $\xi$ using Eqn
(\ref{qpt17}). All the physical quantities can be expressed only in terms
of $x_{\rm c}$ in a certain segment $[x_1,x_2]$ that contains the point
$x_{\rm c}$.

Now, let the temperature $T\neq 0$. We move in the $(x,T\,)$ plane along a
horizontal line $T\neq 0$ (Fig. \ref{Qpt1}, line $aa$) in the
$x\rightarrow x_{\rm c}$ direction. At a certain value of $\delta x$, the
length $\xi _\varphi$, which changes according to Eqns (\ref{qpt16}) and
(\ref{qpt17}), reaches its maximum value $\hbar /T$ [which is determined
by inequality (\ref{qpt15})]. At lower values of $\delta x$, Eqn
(\ref{qpt17}) does not hold on line $aa$, because $\xi$ increases
according to Eqn (\ref{qpt16}) and $\xi _\varphi$ remains equal to its
maximum value $\hbar/T$. The parameters $\xi$ and $\xi _\varphi$ become
mutually independent. The region of this independence is called the
critical region. If we now move toward the transition inside the critical
region (e.g., along line $bx_{\rm c})$, both parameters still diverge at
the transition. However, the divergence of one of them is controlled by
$\delta x$, $\xi \propto \delta x^{\,-\nu }$, and the divergence of the
other is controlled by temperature, $\xi _\varphi =\hbar /T$.

The name of the region derives from the fact that, for a quantum
transition, the critical region is considered to be the region where
quantum fluctuations play an essential role in mixing the two-phase
states. As can be seen from Fig. \ref{Spector2}, this occurs only under
the condition
\begin{equation}\label{16}
|\varepsilon_{02}-\varepsilon_{01}|>T.
\end{equation}
As the temperature decreases, the range of the control parameter $\delta
x$ where condition (\ref{16}) is satisfied narrows. As a result, the shape
of the critical region is unusual: it widens when moving from the
transition.

Inside the critical region, the quantity $\xi _\varphi =\hbar /T$ can be
associated with a real length. The quantity
\begin{equation}\label{qpt18}                   
 L_\varphi\propto(\hbar/T)^{1/z},
\end{equation}
has the required properties: it is equivalent to the parameter
$\xi_\varphi=\hbar/T$ and, according to Eqn (\ref{qpt17}), has the
dimension of length. The physical meaning of $L_\varphi$ can be understood
from the following considerations. Because of a finite temperature, the
quantum problem acquires a characteristic energy $T$ that separates
classical and quantum fluctuations (see Fig. \ref{Spector1}b). Quantum
fluctuations are fluctuations with energies $\hbar\omega _\varphi>T$. The
spatial size of these fluctuations $l_\varphi\propto1/\omega _\varphi$ is
bounded above because frequency $\omega _\varphi >T/\hbar$ is bounded
below. The length $L_\varphi$ is the upper boundary of the size of quantum
fluctuations. Therefore, $L_\varphi$ is often called the dephasing length,
i.e., the length beginning from which coherence in a system of electrons
is broken.

The appearance of two independent parameters in the critical region of a
quantum phase transition is caused by inequality (\ref{qpt15}). Therefore,
a scaling description in the critical region of a quantum phase transition
is therefore called a finite-size scaling. An imaginary thermodynamic
system is thought to exist in a hyperstrip in a ${\cal D}$-dimensional
space with $d$ variables ranging from 0 to $\infty$ and $z$ variables
ranging from 0 to $(\hbar /T\,)^{1/z}$.

Inside the critical region, $L_\varphi<\xi$, and, along its boundary,
\begin{equation}\label{qpt19}                       
 L_\varphi=\xi.
\end{equation}
The lengths $L_\varphi$  and $\xi$ are determined by Eqns (\ref{qpt16})
and (\ref{qpt17}) up to a multiplicative constant; however, Eqns
(\ref{qpt16}) and (\ref{qpt17}) rigidly fix a power relation between these
variables. Therefore, the equation for the critical-region boundaries has
the form
\begin{equation}\label{qpt20}                       
 T=C(\delta x)^{\nu z},
\end{equation}
where the constant $C$ can be different on either side of the transition
in general.

Because the critical region has two independent parameters, the scaling
expressions for physical quantities become more complex. We only present
and discuss expressions for the electric conductivity and resistivity,
\begin{equation}\label{qpt21}
 \sigma=\xi^{2-d}F(L_\varphi/\xi),\qquad
 \rho=\xi^{d-2}F_1(L_\varphi/\xi),\qquad F_1(u)=1/F(u),
\end{equation}
where $F(u)$ is an arbitrary function. The exponent of the first factor in
Eqn (\ref{qpt21}) is specified by how the length enters the expressions
for conductivity at various dimensions $d$. As can be seen from Fig.
\ref{Qpt1}, this factor is determined by the $x$-component of the distance
from the transition. The $L_\varphi /\xi$ ratio is dimensionless;
therefore, scaling rules can be used to regard $L_\varphi /\xi$ as an
argument of an arbitrary function. This relation depends on both $T$
(i.e., the $y$-component of the distance from an image point to the
transition in the phase plane) and the distance from this point to the
critical-region boundary along the $x$-axis.

As an argument, we can also take any power of the $L_\varphi /\xi$ ratio;
as a result, we write the argument in various forms, such as
\begin{equation}\label{argument}
 \frac{L_\varphi}{\xi},\qquad\frac{\hbar/T}{\xi_\varphi},\qquad
 \frac{\hbar/T}{\xi^z},\qquad\frac{\hbar(\delta x)^{z\nu}}{T}\quad
 \mbox{ or }\quad\frac{\delta x}{(T/\hbar)^{1/z\nu}}.
\end{equation}
These various forms elucidate various aspects of the physical meaning of
this ratio. In the last two forms, the argument is dimensional and the
dimensionality of the control parameter is not specified a priori. These
forms show how temperature enters the argument of the scaling function.

\begin{figure}
\includegraphics{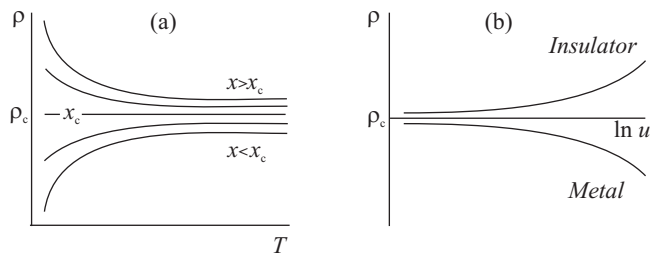}
\caption{(�) Temperature dependence of the resistivity $\rho(T)$ at
various values of $x$ with the horizontal separatrix $x=x_c$. (b)
Reduction of the all the $\rho(x,T)$ curves to two curves (\ref{qpt21}) of
the dependence of the resistivity $\rho$ scaling variable
(\ref{argument})}
 \label{rho2-1}
\end{figure}

In conclusion of this section, as an example, we present an implication of
Eqn (\ref{qpt21}) for the resistivity of a two-dimensional system. If a
quantum phase transition induced by electron localization occurs in a
two-dimensional system and manifests itself in the transport properties of
this system, we can write
\begin{equation}\label{2D-R}
 d=2:\qquad \rho(x=x_c)\equiv1/F(0),\qquad\mbox{i.e.}\qquad
 \rho(x_c,T)=\mbox{const}=\rho_c.
\end{equation}
This implication (a horizontal separatrix in the set of temperature
dependences at various values of $x$) is schematically shown in Fig.
\ref{rho2-1}a. We note that it is valid only under assumption
(\ref{qpt16}), which indicates that the correlation length $\xi$ is
temperature independent. In general, $\rho(x_c,T)$ may have a finite slope
at the point $T=0$ (cf. below, section 5.3 and Fig. 13).

Usually, with $\delta x$ replaced by the modulus $|\delta x|$, the
$\rho(u)$ curve is represented in the form of two branches as a function
of $\ln u$ (Fig. 5b). The values of $z\nu$ are then chosen such that these
curves fit all the experimental points obtained at various temperatures.

\section{3. Flow diagrams for metal--insulator transitions}

So far, we have not specified the type of a quantum phase transition.
Hereafter, we speak about transitions related to a change in the electron
localization, which, in turn, is caused by the degree of disorder in the
system. The general theory of quantum phase transitions initially assumes
that a control parameter affects the interparticle interaction and that a
disorder is not more then a perturbative factor in the initial scheme of
quantum phase transitions. Therefore, the applicability of this scheme to
metal--insulator transitions, in which the degree of disorder is the main
factor and the interaction is a secondary factor, is not obvious a priori.

The fundamental difference between an insulator and a metal is that the
electronic states at the Fermi level are localized in an insulator and
delocalized in a metal. If an insulator is transformed into a metal due to
a change in a certain parameter, the properties of the wave functions at
the Fermi level change. The main physical property that is radically
different in materials of these two types is the conductivity, i.e., the
possibility of carrying an electric current at an arbitrarily weak
electric field. This gives a `yes--no' type signature: the conductivity is
either zero, $\sigma =0$, or nonzero, though it may be arbitrary small.
However, at a finite temperature $T\neq0$, an insulator also carries a
current owing to hopping conductivity. Therefore, the definition of an
insulator given above is only related to the temperature $T=0$, and the
concept of a metal--insulator transition makes sense only at $T=0$.

That the conductivity $\sigma$ is not a function of the state (because it
is realized only under nonequilibrium conditions) is an additional
argument against the applicability of the general theory. However,
Thouless \cite{Thouless} noted that the transport properties can be used
to characterize the decrease in the wave function of an electron placed at
the center of an $L^d$ cube in a $d$-dimensional space when it moves
toward the edges of the cube. Therefore, in analyzing the behavior of
conductance or conductivity near the transition, we actually follow the
evolution of wave functions.

\begin{figure}[t]
\includegraphics{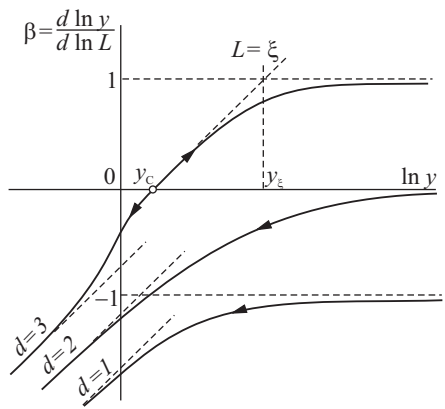}
\caption{Universal $\beta(\ln y)$ functions for different dimentionalities
\cite{b4}.} \label{fmit1}
\end{figure}

Thus, we have every reason to apply the general theory of quantum phase
transitions to metal--insulator transitions. However, the first successful
version of a theoretical description of metal--insulator quantum
transitions \cite{b4} was based on the renormalization group theory
\cite{FiReview} borrowed from quantum field theory. The essence of its
conclusions, related to systems of noninteracting electrons in a random
potential, is illustrated in Fig. \ref{fmit1}, which shows the logarithmic
derivative of the conductance of a sample with respect to the size $L$,
 $$\beta{=}{\frac{\textstyle d\ln y}{\textstyle d\ln
L}}{=}\frac{\textstyle L}{\textstyle y}\frac{\textstyle dy}{\textstyle
dL}$$
 \emph{at the temperature T=0} as a function of the conductance $y$,

\begin{equation}\label{Scaling}                                 
  \frac{\textstyle d\ln y}{\textstyle d\ln L}=\beta_d(\ln y).
\end{equation}

The $\beta_d(\ln y)$ curves describe the universal laws of a change in the
conductance of the system when its sizes change. The behavior of the
system of noninteracting electrons in any material is described by a part
of the curve $\beta_d$ for the corresponding dimension $d$. The
interpretation and method of using these curves are described in detail in
review \cite{LeeR} and book \cite{Gant}.

For metal--insulator quantum phase transitions, the theory \cite{b4} plays
the role of an existence theorem. The shape of the $\beta_d(\ln y)$ curves
and their position in the $(y,\beta)$ plane indicate that these
transitions are absent in one- and two-dimensional systems of
noninteracting electrons and that in three-dimensional systems, such a
transition can occur, with the conductivity of the material changing
continuously during this transition. This means that sufficiently large
one- and two-dimensional samples should be always insulators at absolute
zero whereas a three-dimensional material can be either an insulator or a
conductor. Below, we discuss this feature in detail. Here, we note that
the $\beta_d(\ln y)$ curves are called differential flow diagrams or
Gell-Mann--Low curves.

On the one hand, the theory in \cite{Sachdev,QuantRev,Vojta} is more
specialized than that in \cite{b4} in some respects, because it only
describes the vicinity of a quantum transition. On the other hand, the
theory in \cite{Sachdev,QuantRev,Vojta} is more universal, because its
applicability is not limited by either interaction or a strong magnetic
field. Therefore, it is of interest to compare the conclusions of these
theories for the systems to which they both can be applied.

\section{4. Three-dimensional electron gas}

\begin{figure}[h]
\includegraphics{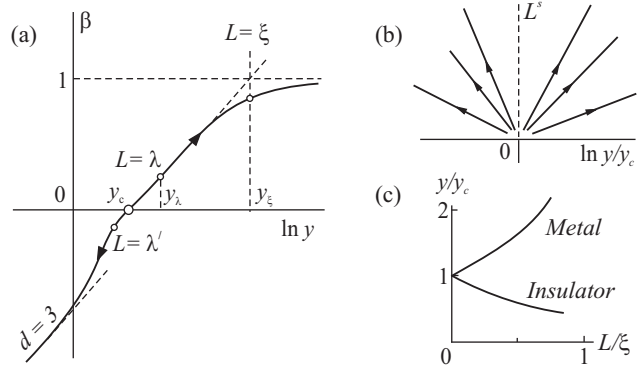}
\caption{Noninteracting 3D electron gas: (a) differential scaling diagram
from \cite{b4} (see also Fig. \ref{fmit1}); the intersection of the curve
$\beta_3(\ln y)$ with the abscissa axis $\beta=0$ means existence of a
quantum phase transition; (b) flow diagram for a noninteracting 3D
electron gas in the vicinity of the transition point; (c) two universal
curves of the dependence of the conductance of the 3D system on its size
$L$ obtained by scaling} \label{3Dqfd}
\end{figure}

The possibility of a phase transition in a three-dimensional material
follows from the fact that the $\beta_3(\ln y)$ curve intersects the
abscissa axis $\beta =0$. A certain point on the $\beta _3(\ln y)$ curve
(an image point) corresponds to the transport properties of a sample of
size $L$ made from a certain material (Fig. \ref{3Dqfd}a). If this point
is located in the lower half-plane $\beta <0$ the material is an
insulator, and the image point shifts to the left as $L$ increases; as a
result, the conductance of the sample becomes exponentially small. If this
point is located in the upper half-plane $\beta >0$ the image point shifts
to the right; as a result, the material is a metal, and its conductance
$y$ increases with the size $L$.

The curves in Figs \ref{fmit1} and \ref{3Dqfd}a are scarcely adapted for a
direct comparison with experimental data because of a special choice of
their coordinate axes. This disadvantage can be partly corrected by
integrating the equation

\begin{equation}\label{3Scaling}
  \frac{\textstyle d\ln y}{\textstyle \beta_3(\ln y)}=d\ln L.
\end{equation}
for three-dimensional (3D) systems. Since the left-hand side of
differential equation (\ref{3Scaling}) becomes infinite at the point
$y=y_{\rm c}$, the curves corresponding to different integration constants
decompose into two families: one of them corresponds to an insulator
region and the other to a metallic region. To demonstrate this by the
simplest way, we restrict ourselves to the immediate vicinity of point
$y_{\rm c}$, in which the $\beta_3(\ln y)$ curve can be approximated by a
straight line,
\begin{equation}\label{tr17}
 \frac{d\ln y}{d\ln L}=s\ln\frac{y}{y_c}
\end{equation}
where $s$ is the slope of the line with respect to the abscissa axis.
Correspondingly, Eqn (\ref{3Scaling}) can be replaced by linear equation
(\ref{tr17}). The general solution of linear differential equation
(\ref{tr17}) is given by
\begin{equation}\label{tr18}                        
   \ln\frac{y}{y_c}=
   \left(\frac L\lambda\right)^{\textstyle s}\ln\frac{y_\lambda}{y_c},
\end{equation}
where $\lambda$  plays the role of the initial condition that fixes the
initial point in the $\beta _3(\ln y)$ curve; for example, it can be the
length $k_{\rm F}^{\,-1}$ determined by the electron concentration. On the
metallic side, $k_{\rm F}^{\,-1}$ is equal to the minimum free path length
$l_{\rm min}$. The conductance $y_\lambda$  corresponds to the point
$\lambda$.

Figure \ref{3Dqfd}b shows particular solutions of Eqn (\ref{tr17}). If
$y_\lambda>y_{\rm c}$, we have $\ln y_\lambda/y_{\rm c}>0$, the solution
is located in the right quadrant, and the conductance $y$ increases with
$L$ (metal). If $y_\lambda <y_{\rm c}$, we have $\ln y_\lambda /y_{\rm
c}<0$, the solution is located in the left quadrant, and the conductance
$y$ decreases as $L$ increases (insulator). The set of curves in Fig.
\ref{3Dqfd}b is called a flow diagram. The arrows in this diagram show the
direction of the image point motion when the system size increases. The
flow lines corresponding to the particular solutions of Eqn (\ref{tr17})
fill both upper quadrants of the $(\ln y_\lambda /y_{\rm c},L^s)$ plane.
The boundary between them, the $\ln y_\lambda /y_{\rm c}=0$ axis, is
called the separatrix; in Fig. \ref{3Dqfd}b, it is indicated by a dashed
line.

We extend straight line (\ref{tr17}) in Fig. \ref{3Dqfd}a  to its
intersection with the asymptote $\beta =1$; as a result, we approximate
the $\beta_3(y)$ curve in the upper half-plane $\beta >0$ by a broken line
consisting of segments of two straight lines. The size $L$ which brings
the image point to the intersection point is called the correlation length
$\xi$. From Eqn (\ref{tr18}), we have
\begin{equation}\label{tr20}
  \xi=\lambda \left(s\ln\frac{y_\lambda}{y_c}\right)^{-1/s}.
\end{equation}
With Eqn(\ref{tr20}), we can rewrite general solution (\ref{tr18}) as
\begin{equation}\label{tr18a}                       
   \ln\frac{y}{y_c}=\left(\frac L\xi\right)^s
\end{equation}
normalize the size $L$ by the length $\xi$ (which is specific for every
material), and reduce each of the two families in Fig. \ref{3Dqfd}b to one
scaling curve (Fig. \ref{3Dqfd}c).

We make several remarks.

First, Eqns (\ref{tr17})--(\ref{tr18a}) involve a parameter $s$, which is
unknown a priori. However, in the case of a noninteracting 3D-gas, this
parameter is either equal to unity or very close to it \cite{LeeR}.
Because $s$ enters all formulas as a factor or exponent, this parameter
can be dropped.

Second, the flow lines in Fig. \ref{3Dqfd}b are straight only because we
integrated the Gell-Mann--Low equation only in the small vicinity of the
transition point. If we integrated the $\beta_3(\ln y)$ function over a
wide range of its argument $\ln y$ using some model representation, we
would obviously obtain curved flow lines.

\begin{figure}[b]
\includegraphics{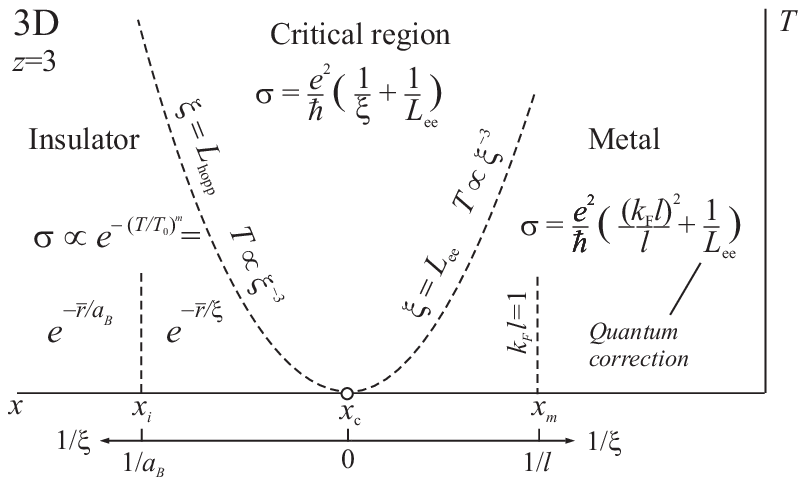}
\caption{Vicinity of a metal--insulator transition in a noninteracting 3D
electron gas in the ($x,T$) phase plane} \label{Qpt3}
\end{figure}

The third note concerns the role of temperature. The parameter that
specifies system motion along the flow lines indicated by arrows in Fig.
\ref{3Dqfd}b is \emph{the size}. If the temperature is taken to be zero
(as was assumed until now), this parameter is the sample size $L$.
However, we can initially suppose that the system is large, as this is
done in the theory of quantum phase transitions
\cite{Sachdev,QuantRev,Vojta}. In this case, the limiting size is
considered to be the dephasing length $L_\varphi$, i.e., the size at which
quantum coherence in an electron system is realized. The image point then
moves along flow lines in the direction indicated by arrows as the
temperature decreases. Because $L_\varphi$ depends on the temperature and
becomes infinite as $T\rightarrow 0$, this procedure introduces the
temperature into the set of physical quantities related to the flow
diagram. However, we do not study this relation; based on the `existence
theorem' of a quantum transition in a 3D space, we turn to the $(x,T\,)$
phase plane to construct the critical region (see Fig. \ref{Qpt3} and
compare it with Fig. \ref{Qpt1}).

Let small values of the control parameter $x$ correspond to metallic
states. Then, at small $x$ on the abscissa axis, the Drude formula $\sigma
=\sigma _3=ne^{\,2}l/\hbar k_{\rm F}$ holds, and, at small $x$ and finite
$T$, the quantum correction
\begin{equation}\label{sigma}
  \sigma=\sigma_3+\frac{e^2}{\hbar}L_{ee}^{-1},\qquad
  L_{ee}=\sqrt{\hbar D/T}.
\end{equation}
is added to $\sigma _3$ (here $D$ is the diffusion coefficient). The
diffusion length in Eqn (\ref{sigma}) is taken to be the length $L_{\rm
ee}$ determined by the Aronov--Altshuler effect \cite{AA}, i.e., by the
electron--electron interaction. In other words, we assume that dephasing
is caused by internal processes that occur in the electron gas in the time
\begin{equation}\label{tau_ee}
\tau_{ee}=\hbar/T,
\end{equation}
without external impacts, such as the electron--phonon interaction (see
analogous discussion of the physical meaning of $L_\varphi$ in review
\cite{QuantRev}, p. 324).

At any control parameter $x$, another special point $x_{\rm m}$ (the Mott
limit) exists on the right of the transition point $x_{\rm c}$. At this
special point, we have
\begin{equation}\label{sigma2}
  \sigma(x_m,T=0)=\sigma_{\rm Mott}=\frac{e^2}{\hbar}k_F=
  \frac{e^2}{\hbar}\frac1l.
\end{equation}

In the range between $x_{\rm c}$ and $x_{\rm m}$, the Drude formula is
invalid: describing conductivity with this formula implies introducing a
free path length $l$ that is shorter than the de Broglie wavelength. In
this range at $T=0$, $\sigma$ is expressed through the correlation length
$\xi$; this corresponds to the general scheme of the theory of quantum
phase transitions. As a result, the conductivity along the abscissa axis
is expressed as
\begin{equation}\label{sigma3}                              
 \sigma(T=0)=\left(\frac{e^2}{\hbar}\right){\times}\left\{
 \begin{array}{lc}
   0 &\qquad x\geqslant x_c, \\
   1/\xi &\qquad x_c\geqslant x\geqslant x_m, \\
   (k_Fl)^2/l &\qquad x\leqslant x_m. \
 \end{array}\right.
\end{equation}
To match the last two expressions at $x=x_{\rm m}$, it suffices to set
$\xi(x_{\rm m})=l$.

Thus, to describe conductivity in the metallic region, we introduced two
parameters, $\xi$ and $L_{\rm ee}$, which have the dimension of length and
the properties required for the critical region (they are mutually
independent and diverge at the point $x_{\rm c})$. Following \cite{Imry},
we construct the following interpolation function in the critical region:
\begin{equation}\label{sigmaCrit}
  \sigma=\frac{e^2}{\hbar}\left(\frac1\xi+\frac{1}{L_{ee}}\right)=
  \frac{e^2}{\hbar}\frac1\xi F\left(\frac{L_{ee}}{\xi}\right)
  \qquad F(u)=(1+1/u).
\end{equation}
Expression (\ref{sigmaCrit}) matches Eqn (\ref{sigma}) on the straight
line $x=x_{\rm m}$ and gives the correct values of conductivity in the
$x_{\rm c}<x\leqslant x_{\rm m}$ segment at $T=0$. The second form of the
function demonstrates that function (\ref{sigmaCrit}) satisfies general
relation (\ref{qpt21}).

We now move from right to left along the line $T={\rm const}$ (Fig.
\ref{Qpt1}, line $aa$). As long as the quantum correction is relatively
small, electron diffusion occurs via scattering by impurities; i.e., it is
controlled by the first term in Eqn (\ref{sigma}). Therefore, the
diffusion coefficient $D$ in $L_{\rm ee}$ is temperature independent.
However, when we enter the critical region, $\sigma_3$ transforms into
$(e^{\,2}/\hbar )\xi ^{\,-1}$ and begins to decrease rapidly. Under these
conditions, $D$ ceases to be a constant: diffusion is likely to be caused
by the electromagnetic-field fluctuations that determine $L_{\rm ee}$; as
a result, this diffusion becomes temperature independent. We can then
write a set of equations for the $\sigma (T\,)$ and $D(T\,)$ functions
with the Einstein relation
\begin{equation}\label{Einst}               
 \left\{
  \begin{array}{l}
    \sigma={\displaystyle\frac{e^2}{\hbar}
    \left(\frac1\xi+\sqrt{\frac{T}{\hbar D}}\right)}, \\
    \sigma=e^2g_FD \rule{0pt}{5mm} \
  \end{array}
 \right.
\end{equation}
used as a second equation. Here $g_{\rm F}$ is the density of states at
the Fermi level.

We eliminate $D$ from two equations (\ref{Einst}), use that $1/\xi \ll
1/L_{\rm ee}$ near the transition, and obtain the temperature dependence
of the conductivity on the right-hand side of the critical region
\cite{Imry,AA-Lett},
\begin{equation}\label{t3}
  \sigma(T)=\frac{e^2}{\hbar}\left(\frac{1}{\xi}+(Tg_F)^{1/3}\right)
  \equiv\alpha+\beta T^{1/3}.
\end{equation}
This means that inside the critical region, $L_{\rm ee}$ is given by
\begin{equation}\label{Lee-crit}        
  L_{ee}=(Tg_F)^{-1/3}.
\end{equation}
rather than being determined by Eqn (\ref{sigma}). Exactly at the
transition, we have $\xi =\infty$ and $\alpha =0$.

\begin{figure}[t]               
\includegraphics{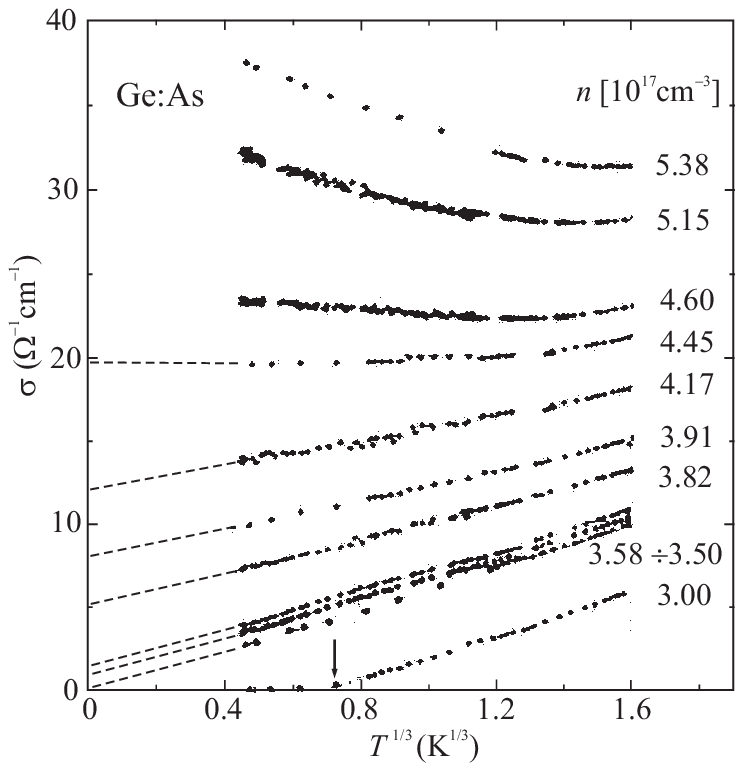}\hspace{2cm}\includegraphics{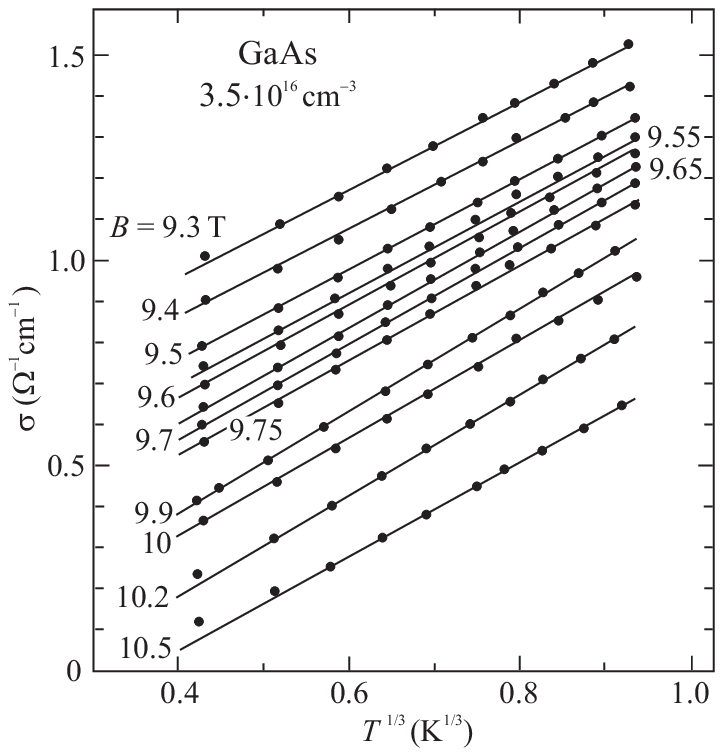}
\caption{(a) Temperature dependence of the conductivity of Ge:As samples
with various level of doping in the region of a metal--insulator
transition (from \cite{E}); the critical concentration is determined by
the extrapolation of the experimental data to $T=0$. b) Temperature
dependence of the conductivity of a GaAs sample in various magnetic fields
in the region of a metal--insulator transition (from \cite{C}); the
critical field determined by extrapolation is 9.78 T.}
 \label{MITexp}
 \end{figure}

Temperature dependence  (\ref{t3}) was repeatedly detected in experiments
performed on a variety of materials \cite{A,B,C,D}. Figure \ref{MITexp}
shows two examples in which the control parameter is represented by the
electron concentration and magnetic field.

 Using condition (\ref{qpt19}) and Eqn (\ref{t3}), we can write the relation
\begin{equation}\label{CubicPar}
  T\propto(g_F\xi^3)^{-1}.
\end{equation}
for the right boundary of the critical region. If the relation $\xi \propto
(\delta x)^{-1}$ holds (i.e., if $\nu =1$, as is assumed in Fig. \ref{Qpt3}),
the boundary of the critical region is represented by a cubic parabola
$T\propto (\delta x)^3$. In the general case, we have $\nu \neq1$ and
\begin{equation}\label{nuCubicPar}
  T\propto(\delta x)^{3\nu}/g_F.
\end{equation}
A comparison with Eqn (\ref{qpt20}) demonstrates that in the metal--insulator
transition in a 3D system of noninteracting electrons, the dynamic critical
index is
\begin{equation}\label{qpt24}                   
 z=3.
\end{equation}

Below curve (\ref{nuCubicPar}) in the $x_{\rm c}>x>x_{\rm m}$ segment, the
conductivity is also described by Eqn (\ref{sigmaCrit}); however, $L_{\rm ee}$
enters Eqn (\ref{sigmaCrit}) in the initial form (\ref{sigma}) with the
diffusion coefficient $D={\rm const}$. Therefore, the temperature dependence of
the conductivity in this region should have the form $\sigma (T\,)=\alpha+\beta
T^{\,1/2}$.

Up to this point, we have discussed the right-hand side of the phase diagram.
The left-hand side of the phase diagram, i.e., the insulator region (where
hopping conductivity takes place), also has two characteristic lengths. First,
there is the decay length $\xi$ of the localized-state wave functions, $\psi
\propto \exp {(-r/\xi )}$. Far from the transition, $\xi$ decreases to the Bohr
radius $a_{\rm B}=\kappa \hbar ^{\,2}/(m^{\,*}e^{\,2})$ (where $\kappa$ is the
dielectric constant and $m^{\,*}$ is the effective electron mass); at the
transition, it diverges because the electrons become delocalized. Second, there
is the average hopping distance $\overline r$. If the hopping conductivity is
described by the Mott law, the average hopping distance is \cite{Gant}
\begin{equation}\label{hop}
 \overline{r}=(\xi/g_BT)^{1/4}.
\end{equation}
The lengths $\overline r$ and $\xi$ cannot be used as two independent lengths
in the critical region because they are connected by relation (\ref{hop}).
However, using $L_{\rm ee}$ determined by Eqn (\ref{Lee-crit}), we can rewrite
Eqn (\ref{hop}) as
\begin{equation}\label{hop1}                        
 \overline{r}=(\xi L_{ee}^3)^{1/4},
\end{equation}
We take into account that expression (\ref{Lee-crit}) for $L_{\rm ee}$ does not
contain the kinetic characteristics of the electron gas and consider $L_{\rm
ee}$ the dephasing length over the entire critical region, including its
left-hand side, above the $x_{\rm i}>x>x_{\rm c}$ segment on the abscissa axis.
Equation (\ref{nuCubicPar}) then determines both the right and left boundaries
of the critical region. Equations (\ref{hop}) and (\ref{hop1}) demonstrate that
$\overline {r}=\xi =L_{\rm ee}$ at the left boundary.

The conclusion that $L_{\rm ee}$ can be considered to be the dephasing length
over the entire critical region is supported experimentally: as can be seen
from Fig. \ref{MITexp}, the $\sigma (T\,)$ temperature dependence straighten in
terms of the $(T^{\,1/3},\sigma)$ axes, not only on the right-hand side of the
critical region but also at values of the control parameter $x<x_{\rm c}$
(e.g., at the electron concentration $n=3\times10^{17}\,$cm$^{-3}$ in Fig.
\ref{MITexp}). However, the free term $\alpha$ in Eqn (\ref{t3}) becomes
negative. This means that we should either suppose that correlation length
$\xi$ is negative in the insulator region or (which is formally preferred but
is essentially the same) should replace interpolation formula (\ref{t3}) on the
left-hand side of the critical region by the formula
\begin{equation}\label{t3-1}
  \sigma(T)=\frac{e^2}{\hbar}\left(-\frac{1}{\xi}+\frac{1}{L_{ee}}\right).
\end{equation}
As follows from Eqns (\ref{hop1}) and (\ref{t3-1}), we have $\sigma =0$ along
the left boundary of the critical region, which means that the conductivity
along this boundary is determined up to an exponentially small hopping
conductivity.

The hopping conduction mechanism is still operative below the lower boundary of
the critical region over the $x_{\rm i}>x>x_{\rm c}$ segment, and the
wavefunction decay length is $\xi \gg a_{\rm B}$ rather than $a_{\rm B}$
\cite{Castner}. Therefore, $a_{\rm B}$ does not enter the expression for the
hopping conductivity, and $\sigma$  is expressed in terms of the correlation
length $\xi$ (see Fig. \ref{Qpt3}):
 $$\sigma\propto\exp(-\overline{r}/\xi).$$

The $(x,T\,)$ phase diagram in Fig. \ref{Qpt3} accumulates the results of the
long-term experimental studies of the low-temperature transport properties of
conducting systems, namely, the quantum corrections to metallic conductivity,
the evolution of electronic spectra during metal--insulator transitions, the
temperature dependence of conductivity in the vicinity of the transitions, and
hopping conductivity. In essence, this phase diagram was plotted irrespective
of the theory of quantum phase transitions \cite{Sachdev,QuantRev,Vojta} in
order to reveal the compatibility of all experimental data. Nevertheless, the
considerations given above demonstrate that this diagram is absolutely adequate
for the concepts following from this theory.

\section{5. Two-dimensional electron gas.}

{\bf 5.1 Gas of noninteracting electrons}

\begin{figure}[b]
\includegraphics{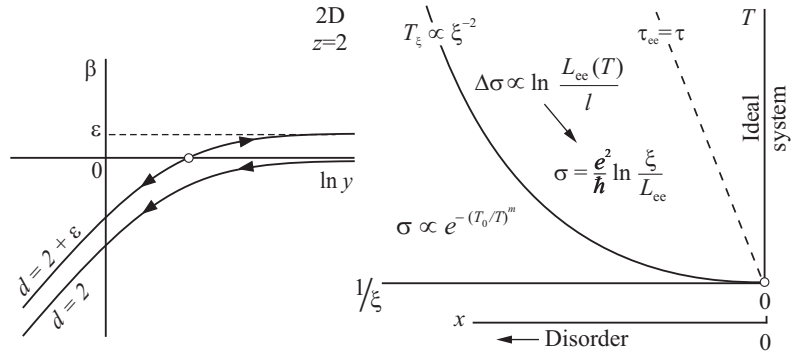}
\caption{Noninteracting 2D electron gas: (a) differential flow lines for
systems with dimension $d=2$ taken from the scaling diagram in \cite{b4}
(see also Fig. \ref{fmit1}) and $d=2+\varepsilon$ (see text), and (b)
crossover from the logarithmic temperature dependence of conductivity to
its exponential dependence, which can be treated as the boundary of the
critical region of a virtual quantum transition  (see text)}
 \label{Qpt4}
\end{figure}

We now pass to 2D systems. According to the theory by Abrahams et al. [7], a 2D
system $(d=2)$ of noninteracting electrons is always an insulator in the sense
that localization should inevitably occur in a sufficiently large sample,
$L>\xi$, at a sufficiently low temperature, $T<T_\xi$, and an arbitrarily small
disorder. This statement follows from the fact that flow line $d=2$ in Fig.
\ref{fmit1} asymptotically approaches the $\beta =0$ axis and does not
intersect it (this line is shown in Fig. \ref{Qpt4}a). However, in weakly
disordered films, the correlation length $\xi$, which bounds the size $L$
below, can be unrealistically large, and the temperature $T_\xi$ of the
crossover from the region with a logarithmic quantum correction to the
conductivity to the region that is characterized by an exponential temperature
dependence is, in contrast, too low. These films are called metallic films.

Estimates of $\xi$ and $T_\xi$ can be obtained from the assumption \cite{LKh}
that the logarithmically diverging quantum correction $\Delta \sigma$ in the
conductivity
 \begin{equation}\label{01}                             
  \sigma=\sigma_2+\Delta\sigma=(e^2/\hbar)(k_Fl)
  -(e^2/\hbar)\ln(L_{ee}/l)
 \end{equation}
is of the order of the classical conductivity $\sigma _2$, and hence $\sigma
\approx 0$ and
 \begin{equation}\label{sigma0}
  \ln(L_{ee}/l)\approx k_Fl.
 \end{equation}

The diffusion length $L_{\rm ee}$ over which electron dephasing occurs is
determined by Eqn (\ref{sigma}). The value of $L_{\rm ee}$ determined from
Eqn (\ref{sigma0}) is defined as the correlation length $\xi$,
\begin{equation}\label{2a}
 \xi=l\exp(k_Fl),
 \end{equation}
and the temperature
\begin{equation}\label{5}                   
 T_\xi=D\hbar/\xi^2,
 \end{equation}
determined from the relation $L_{\rm ee}=\xi$ using Eqn (\ref{sigma}) is called
the crossover temperature. Of course, the conductivity does not vanish at
$T=T_\xi$; however, the theory of weak localization is obviously not valid at
this temperature, and the resistivity should begin to increase exponentially
with decreasing the temperature at $T<T_\xi$ in samples of size $L>\xi$.

We now plot function (\ref{5}) and, for convenience, direct the $1/\xi$ axis to
the left (see Fig. \ref{Qpt4}). As a result, this diagram can be conveniently
compared with the diagram for a 3D system shown in Fig. \ref{Qpt3}. We add an
axis and lay off the control parameter $x$ as abscissa; this parameter
characterizes the degree of disorder, with $x=0$ corresponding to an ideal
system in which a disorder is absent.

It is easily seen that the $T(1/\xi )$ curve in Fig. \ref{Qpt4} represents the
left-hand side of the phase diagram in the vicinity of a metal--insulator
transition in a 3D material (cf. Fig. \ref{Qpt3}) whose phase transition point
is located at the edge of the diagram, at the origin $(T=x=0)$. Curve (\ref{5})
is then the left boundary of the critical region, and the dynamic critical
index is
\begin{equation}\label{z=2}                 
 z=2.
\end{equation}

Using Eqns (\ref{qpt24}) and (\ref{z=2}), one can conclude that for
metal--insulator transitions in systems of noninteracting electrons, the
dynamic critical index is equal to the dimension, $z=d$.

Thus, systems with dimension $d=2$ turn out to be boundary systems: a quantum
transition is still present in the $(x,T\,)$ phase plane but is shifted toward
its corner, to the unreachable point $x=0$. The boundary properties of 2D
systems can also be found from the flow diagram in Fig. \ref{fmit1}. We imagine
that the dimension $d$ is a continuous parameter and can take not only integer
values. Straight lines $\beta=d-2$ represent the asymptotics of the flow lines
at high values of $y$. Therefore, the flow lines of a system with dimension
$d=2+\epsilon>2$ inevitably cross the abscissa axis $\beta=0$, and such systems
have a metal--insulator transition (Fig. \ref{Qpt4}a). As
$\epsilon\rightarrow0$, the transition point shifts toward high conductance and
goes to infinity.

This interpretation of the curve in Fig. \ref{Qpt4} implies that the domain
over the $T\propto (1/\xi )^2$ parabola is the critical region of the quantum
transition. For 2D systems, only one scaling variable $u$ is retained in Eqn
(\ref{qpt21}) written for the conductivity in the critical region; it is equal
to the ratio of two characteristic lengths,
\begin{equation}\label{qpt27}
 \sigma=F(u)\equiv F(L_\varphi/\xi).
\end{equation}
In this region, however, the conductivity is typically expressed as the
difference between the classical conductivity and the quantum correction,
\begin{equation}\label{qpt28}
 \sigma=\sigma_2-\Delta\sigma=
 \frac{e^2}{\hbar}\left(k_Fl-\ln\frac{L_{ee}}{l}\right).
\end{equation}
To resolve this apparent discrepancy, let us substitute the expression $l=\xi
\exp {(-k_{\rm F}l)}$ from Eqn (\ref{2a}) in the argument of the logarithm in
Eqn (\ref{qpt28}), move the exponent from the logarithm, and obtain
 \begin{equation}\label{30}
 \sigma=\frac{e^2}{\hbar}\ln\frac{\xi}{L_{ee}}.
 \end{equation}
The classical conductivity $\sigma _2$ cancels and the remaining part depends
only on the scaling variable, as it should be in the critical region. Thus, in
this regard, our $(x,T\,)$ diagram also satisfies the requirements of the
theory of quantum phase transitions.

Expression (\ref{30}) holds not in the entire `quasi-critical' region
$T>T_\xi$; moreover, it diverges on the $x=0$ axis. However, near this axis,
the elastic mean free path  $l\rightarrow \infty$  and the standard expression
for $L_{\rm ee}$ lose their meaning because the necessary condition $\tau _{\rm
ee}\gg \tau$ is violated. The boundary of the part of the region where Eqn
(\ref{30}) is invalid, $\tau _{\rm ee}=\tau$, is shown in Fig. \ref{Qpt4} by a
dashed straight line plotted under the assumption that $\tau \propto x^{\,-1}$.

As is shown in Section 5.3, the introduction of interaction can lead to the
appearance of a metal--insulator transition in a 2D system. In terms of the
phase diagram shown in Fig. \ref{Qpt4}, this means that the phase transition
point shifts from the origin to a point $x_{\rm c}\neq0$ on the abscissa axis
due to a certain cause, and a boundary additionally appears between the
critical and metallic regions. Standard expression (\ref{qpt28}), which was
written for the conductivity in the metallic region and was transformed into
Eqn (\ref{30}), was already used for the description of the conductivity in the
critical region. This means that the expression for the conductivity in the
metallic region should be radically different and that we should expect the
appearance of a marginal metal instead of a Fermi liquid. The nature of the
electronic states in this hypothetic metal should be rather peculiar, since
electrons are assumed to be localized when interaction is turned off
\cite{Elihu}.

 {\bf 5.2. Spin--orbit interaction}

The boundary position of 2D systems makes them sensitive to various types of
interaction, e.g., spin--orbit interaction or electron--electron interaction,
which can cause a phase transition.

We first consider the spin--orbit interaction. This case is convenient because
a flow diagram can be plotted using the $(y,\beta)$ axes of the initial diagram
shown in Fig. \ref{fmit1}. At large $y$, the initial flow line $\beta_2(y)$
deviates down from its right asymptote $\beta=0$ because of weak localization,
which results in a decrease in the conductivity (Fig. \ref{Qpt4}a). But the
spin--orbit interaction changes the sign of the quantum correction, i.e.,
changes weak localization into antilocalization. Therefore, the sign of the
derivative on the right-hand side of the $\beta_2(\ln y)$ curve should change:
the curve deviates upward from the asymptote $\beta =0$, goes to the upper
half-plane $\beta >0$, and, hence, inevitably crosses the abscissa axis
$\beta=0$ and reaches the left asymptote.

\begin{figure}[h]                           
\includegraphics{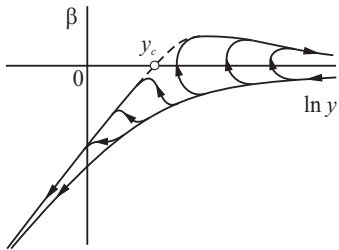}
\caption{Flow diagram for a 2D electron gas with spin-orbit interaction.}
 \label{2Dso}
\end{figure}

Antilocalization was studied in detail both experimentally \cite{Berg} and
theoretically \cite{Lark}. In the 2D case $(d=2)$, the quantum correction to
the conductivity is given by
\begin{equation}\label{so2}
 \Delta\sigma_2\approx
  -(e^2/\hbar)\int\limits_\tau^{\tau_{ee}\rule[-1.5mm]{0pt}{1mm}}
  \frac{dt}{t}\left(\frac32\,e^{-t/\tau_{so}}-\frac12\right),
\end{equation}
where $\tau$ is the time between elastic collisions, $\tau _{\rm so}\ge \tau$
is the time between spin-flip collisions, and $\tau _{\rm ee}=\hbar /T$ is the
dephasing time. The motion of the image point in the flow diagram now depends
on two parameters, the conductance $y$ and time $\tau _{\rm so}$. For various
values of $\tau _{\rm so}$, Fig. \ref{2Dso} shows the family of flow lines
located between two envelope curves. The lower curve is $\beta_2(\ln y)$ from
the diagrams in Figs \ref{fmit1} and \ref{Qpt4}. It can be obtained from
integral (\ref{so2}) if we set $\tau _{\rm so}=\infty$ in the integrand, and
hence the parenthesis in the integrand becomes equal to unity. The upper curve
has the same asymptotics; however, its construction implies that $\tau _{\rm
so}=\tau$. Then, we have $t\gg \tau _{\rm so}$ over the major portion of the
integration range, and the parenthesis is considered to be $-1/2$.

We now take a sample of size $L$ of a material with an intermediate value of
$\tau _{\rm so}$,
 $$\tau\ll\tau_{so}\ll\infty.$$
The size $L$ bounds the diffusion time $t$ of an electron in a 2D sample until
the collision with boundary by the quantity $\tau _L, t<\tau _L\sim \tau
(L/l\,)^2$. Therefore, to describe weak localization in a sample of size $L$,
we must replace the upper integration limit in Eqn (\ref{so2}) with $\tau _L$,
$$\tau_{ee}\rightarrow\tau_L\equiv\tau(L/l)^2.$$
 Let $L$ be first very small $(L\,\\gtrsim1)$ and the temperature $T=0$.
Diffusion processes then have no time to develop; electron interference is
virtually absent; the conductivity is equal to its classical value; and the
image point is on the right on the axis $\beta =0$. As the size $L$ increases
to
 \begin{equation}\label{Lso}
 L_{so}=l\sqrt{\tau_{so}/\tau}
 \end{equation}
spin-flip collisions are insignificant and correction (\ref{so2}) to the
conductivity is negative. The image point moves to the left along the lower
envelope curve, as in Fig. \ref{Qpt4} in the absence of the spin--orbit
interaction. When the size $L$ becomes larger than $L_{\rm so}$, integral
(\ref{so2}) and $\Delta \sigma _2$ change their sign and the image point moves
to the upper envelope curve. If this passage occurs at $y<y_{\rm c}$, the film
becomes an insulator at $L\rightarrow \infty$ in any case. But if the
spin--orbit interaction is strong, $\tau _{\rm so}$ is small and the image
point reaches the upper envelope curve at $y>y_{\rm c}$. Then, the image point
continues to move along the upper envelope curve toward large $y$, and the film
retains its metallic properties as $L\rightarrow \infty$.  Thus, a
metal--insulator transition could be observed in an experiment with the
spin--orbit interaction used as a control parameter. So far, only the
transformation of weak localization into antilocalization has been demonstrated
this way \cite{Berg}.

\vspace{5mm}{\bf 5.3. Gas of interacting electrons}

We have to refine what is meant by the absence or presence of the
electron--electron interaction. The interaction manifests itself differently in
the properties of electron systems; for example, it determines the probability
of electron--electron scattering. In the classical theory of metals,
electron--electron scattering is considered not to contribute to conductivity,
since the total momentum of the electron system and the drift velocity remain
the same. When quantum effects are taken into account, this statement becomes
invalid because the conductance depends on the dephasing length and
electron--electron scattering changes this length. Nevertheless, if the
electron--electron interaction affects the conductance only through scattering,
the electron system is considered noninteracting, because the interaction
affects the conductance as an external action, e.g., scattering by phonons.

Electron--electron scattering is not the only channel of the effect of the
interaction on the conductance. The interaction is thought to be not the bare
Coulomb interaction but the interaction between `dressed' quasiparticles, i.e.,
the screened interaction that depends on the electron density, the diffusion
coefficient of electrons, and (under certain conditions) the sample size or the
quantum coherence length \cite{McMillan}.

This interaction determines the structure of quantum levels and the
ground-state energy, affecting the competition between phases in the vicinity
of a phase transition. Electrons are mainly scattered by a screened impurity
potential, whose properties depend on the effective electron--electron
interaction. This interaction, in turn, depends on the diffusion coefficient
and dephasing length, which results in the complex dependence of the effective
interaction on all external parameters.

The case of spin--orbit interaction considered in Section 5.2 represents an
example of interaction renormalization as the sample size $L$ changes: the
spin--orbit interaction is actually turned on only when $L$ becomes larger than
the value determined from Eqn (\ref{Lso}).

\begin{figure}[b]                       
\includegraphics{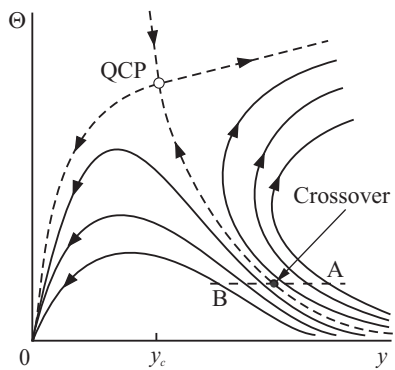}
\caption{Part of a flow diagram plotted for a 2D electron gas with interaction
\cite{Fink} }
 \label{Qpt5}
\end{figure}
As another example, we analyze the picture of possible states in the model with
a multivalley electron spectrum developed by Punnoose and Finkel'stein
\cite{Fink}. In this model, a change in $L$ causes changes in both the
conductance y and interaction $\Theta$. Therefore, an equation for $\Theta$ is
added to Eqn (\ref{Scaling}). The flow diagram is a result of the solution of
these two equations. Figure \ref{Qpt5} shows part of this diagram calculated by
the authors of \cite{Fink} in slightly different coordinate axes. The variables
were changed to facilitate a comparison of this diagram with those shown in
Figs�6 and 10, although the quantitative information contained in the initial
diagram in \cite{Fink} is lost. The abscissa of the flow diagram shown in Fig.
\ref{Qpt5} is the conductance. Thus, the abscissa of all the diagrams in Figs
\ref{fmit1} and \ref{Qpt4} is the same. The ordinate $\Theta$ of the flow
diagram in Fig. \ref{Qpt5} reflects the effective interaction. A noninteracting
electron gas, which was discussed in Section 5.1, corresponds to the straight
line $\Theta =0$. As in all previous diagrams (Figs \ref{fmit1}--\ref{Qpt4}),
the size is the parameter that determines system motion along the flow lines
indicated by arrows. This size is given by the sample size $L$ if $T=0$ or by
the dephasing length $L_\varphi$  (or $L_{\rm ee}$), i.e., the maximum size in
which quantum coherence is retained in an electronic system.

The configurations of the flow lines in Fig. \ref{Qpt5} clearly display the
specific feature of the interaction $\Theta$  that was discussed at the
beginning of this section: \emph{as the size $L$ or $L_\varphi$  changes, the
interaction in the system changes (is renormalized)}, and this change is
different in different flow lines. The diagram in Fig. \ref{Qpt5} is a
two-parameter diagram: two parameters are required to specify the flow line of
the image point. The flow lines (trajectories) in this diagram occupy the
entire half-plane.

We assume that the position of a point in a flow line is determined by
$L_\varphi$, i.e., by the temperature. We assume that we are at point A in a
flow line, that is metallic because it recedes to the region of high
conductivity at $T\rightarrow 0$. We fix the interaction $\Theta$ and vary the
conductance $y$; that is, we vary the degree of disorder. This process is shown
by a dashed straight line in the diagram. When moving along this line, we can
cross the separatrix and reach point B in a flow line that describes an
insulator, because it tends to the point $y=0$ as $T\rightarrow 0$. To specify
the position of a point in line AB, we can use a single parameter, which is
called a control parameter. At another control parameter, the angle of
intersection of the separatrix can be different. For example, we can initiate
the crossover from a metallic to a nonmetallic flow trajectory by changing the
electron concentration and, thus, simultaneously changing the effective
interaction $\Theta$ and the conductance.

\begin{figure}[h]                           
\includegraphics{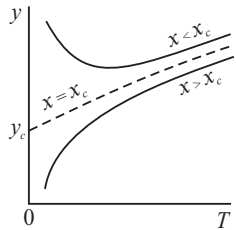}
 \caption{Qualitative scheme for the evolution of the conductance $y(T)$ curves
with a control parameter in the model proposed in \cite{Fink}} \label{y(T)}
\end{figure}

By changing the control parameter, we can pass from one flow trajectory to
another, and, by changing the temperature (size), we can move along a flow
trajectory. But because interaction $\Theta$ can change when any of two
parameters changes, factorization is absent; that is, we cannot suppose that
the length $\xi$ depends only on $\delta x$ and the length $L_\varphi$ depends
only on temperature. The consequence of this `mixing' of variables is clearly
visible in the flow diagram in Fig. \ref{Qpt5}. As we move along the separatrix
toward the quantum critical point (QCP), the conductance decreases and tends to
$y_{\rm c}$. Hence, we can draw the following important conclusion, which is
qualitatively shown in Fig. \ref{y(T)}: the separatrix in the set of
temperature dependence is not a horizontal line, as in Fig. \ref{rho2-1}.

\begin{figure}[t]                           
\includegraphics{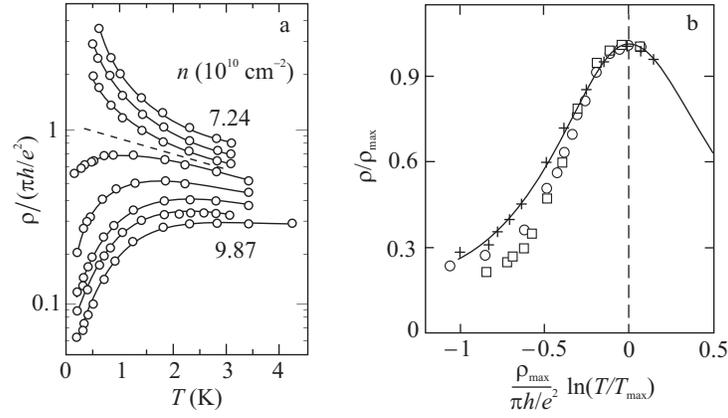}
 \caption{a) Temperature dependence of the resistivity of a 2D gas in a
field-effect silicon transistor in a concentration range containing the
metal--insulator transition \cite{Fink1}. b) Scaling of the three lower curves,
which correspond to the concentrations  9.87 ($\square$), 9.58 ($\bigcirc$),
and 9.14$\cdot10^{10}$\,cm$^{-2}$ ($+$). Theoretical results are shown by a
solid line, \cite{Fink1}} \label{R(T)exp}
\end{figure}

An indication of the presence of a metal--insulator transition in a 2D gas was
first obtained in the inversion layer of a field-effect transistor on a Si
surface \cite{Krav}. The presence of the transition was questioned for a long
time, because is was in conflict with the concepts formulated in \cite{b4} and
because the transition was not reproduced in other materials. However, the
uniqueness of silicon was found to be related to a high electron mobility,
which allows performing experiments at a very low electron density, where the
electron--electron interaction is especially important. With this fact, it was
possible to interpret the experimental data using not the theory in \cite{b4},
which was developed for noninteracting electrons, but in accordance with the
model in \cite{Fink}. As a result, the presence of a metal--insulator
transition was supported and a finite slope of the separatrix, which was
predicted in \cite{Fink}, was obtained (see Fig. \ref{R(T)exp}a borrowed from
\cite{Fink1}.

As is seen from the flow diagram in Fig. 12, the finite slope of the separatrix
in the set of the temperature dependence of $\sigma$ or $R$ of \emph{a system
of 2D electrons} is controlled by the angle at which the separatrix in the flow
diagram approaches the QCP. If the tangent to the separatrix is normal to the
abscissa axis at the QCP due to any specific reason, the separatrix in the set
of temperature dependence has a zero derivative at $T=0$. Thus, the horizontal
position of the separatrix in the set of the temperature dependence of
conductivity in the vicinity of the quantum phase transition results from the
symmetry of the flow diagram of a certain system and is not an inherent
property of all 2D systems. Conversely, a finite slope of the separatrix is not
an indispensable consequence of a two-parameter flow diagram.

An inclined separatrix complicates scaling, i.e., the reduction of measurements
performed along different flow lines to one universal curve by changing the
scales. Nevertheless, the scaling of the resistance $R(T\,)$ data is possible.
Figure \ref{R(T)exp}b shows the scaling carried out in the metallic region of
the transition displayed in Fig. \ref{R(T)exp}a. The three lower curves in Fig.
\ref{R(T)exp}a are replotted in the coordinates $\rho /\rho _{\rm max}$
(instead of $\rho$ ) and $\rho _{\rm max} \ln {(T/T_{\rm max} )}$ (instead of
$T$; here, $\rho _{\rm max}$ is given in dimensionless units), and the values
$\rho _{\rm max}$  and $T_{\rm max}$ correspond to the position of the maximum
in each of the experimental curves, which are seen to merge into one curve and
to coincide with the theoretical curve. This last curve was plotted using the
calculations in \cite{Fink} and the interaction parameter $\Theta$  determined
in the same sample from the magnetoresistance data obtained in a field parallel
to the 2D plane.

An analogous problem of an inclined separatrix is also often encountered during
the experimental processing of the $R_x(T\,)$ temperature dependence ($x$ is a
control parameter) in the vicinity of superconductor--insulator quantum phase
transitions. In Ref. \cite{Scaling}, a procedure was proposed for the
`correction' of the curves via the introduction of the additional linear term
 \begin{equation}\label{Scal}
 R_x(T) \rightarrow R_x(T)-\alpha T,\qquad\alpha=
 \left.\frac{\partial R_{x_c}(T)}{\partial T}\right|_{T=0},
 \end{equation}
into each of them to make the separatrix horizontal and for performing
subsequent standard scaling for 2D systems using scaling variable
(\ref{argument}). The idea of this procedure is to compensate for the slope of
the separatrix in a flow diagram. However, the correctness of this procedure
has not been proved theoretically.

\begin{figure}[h]                       
\includegraphics{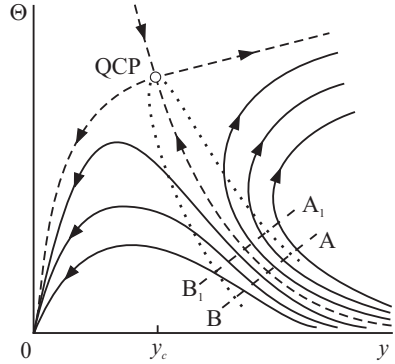}
\caption{Critical region in the flow diagram of a 2D electron gas with
interaction (indicated by a dotted curve) }
 \label{Qpt5cr}
\end{figure}

Formally, the $(x,T\,)$ phase plane also has a meaning for a two-parameter flow
diagram. However, the critical vicinity of the transition can also be directly
plotted in the diagram (see Fig. \ref{Qpt5cr}). It should be noted that the
two-parameter scaling becomes one-parameter in the region adjacent to the
separatrix inside the critical region, where flow trajectories are parallel.
Indeed, as noted above, one parameter is sufficient to specify the position of
a point on line AB, and this parameter does not change when line AB shifts
parallel to the separatrix, e.g., to position A$_1$B$_1$, since the flow lines
are parallel to each other. As we move along the separatrix toward the quantum
critical point, near-separatrix trajectories turn aside alternately, and the
strip in which the flow trajectories are parallel, as well as the critical
region, narrows.

\section{6. Quantum transitions between the different states of a Hall liquid}

States in the plateaus of the quantum Hall effect are the specific phase
states of a 2D electron gas with special transport properties described by
the longitudinal $\sigma _{xx}$ and transverse $\sigma _{xy}$
conductivities
\begin{equation}\label{q201}
  \sigma_{xx} \rightarrow0\quad\mbox{at}\quad T\rightarrow0,\qquad
  \sigma_{xy}=i(e^2/2\pi\hbar),\qquad i=0,1,2,3...\:.
\end{equation}
Such phase states of a 2D electron gas are quantum Hall liquids with
different quantum Hall numbers $i$, which are determined by the values of
the Hall conductivity $\sigma _{xy}$, Eqn (\ref{q201}), in the plateaus,
\begin{equation}\label{q202}
 i=\sigma_{xy}/(e^2/2\pi\hbar),\qquad i=1,2,3...\:.
\end{equation}

\begin{figure}[h]                       
\includegraphics{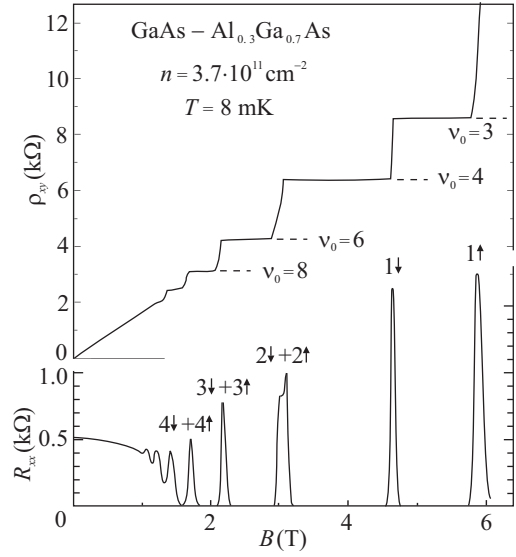}
\caption{The magnetoresistance $R_{xx}$ and the Hall resistivity $\rho_{xy}$
versus the magnetic field $B$ in a GaAs--Al$_x$Ga$_{1-x}$As heterostructure at
$T=8$\,mK \cite{K1}. The electron density in the 2D layer is
$3.7\cdot10^{11}$\,cm$^{-2}$, and the mobility is
$\mu=4.1\cdot10^4$\,cm$^2$/V$\cdot$s.} \label{fq8e}
\end{figure}

The transitions from one plateau to another induced by a change in the
magnetic field or the electron concentration are clearly visible in
experimental curves. As can be seen from Fig. \ref{fq8e}, $\sigma _{xy}$
jumps are accompanied by narrow $\sigma _{xx}$ peaks. These jumps are
quantum phase transitions, and they should fit both theoretical versions
compared in Section 5.

Figure \ref{fq8e} only displays the plateaus of the integer quantum Hall
effect, which is considered below. The integer quantum Hall effect can
also be realized in a noninteracting electron gas; therefore, it can be
described without regard for interaction.

To construct the flow diagram of a 2D system of noninteracting electrons
in a strong magnetic field, we need two conductance components that are
equivalent to conductivity components $\sigma _{xx}$ and $\sigma _{xy}$.
Correspondingly, Eqn (\ref{Scaling}) transforms into the set of two
equations
\begin{equation}\label{301}
 \begin{array}{c}
   \frac{\textstyle d\ln\sigma_{xx}}{\textstyle d\ln L\rule{0pt}{3mm}}
    =f_1(\sigma_{xx},\sigma_{xy}), \\
   \frac{\textstyle d\ln\sigma_{xy}}{\textstyle d\ln L\rule{0pt}{3mm}}
    =f_2(\sigma_{xx},\sigma_{xy})   \rule{0pt}{8mm},
 \end{array}
\end{equation}

\begin{figure}[h]                       
\includegraphics{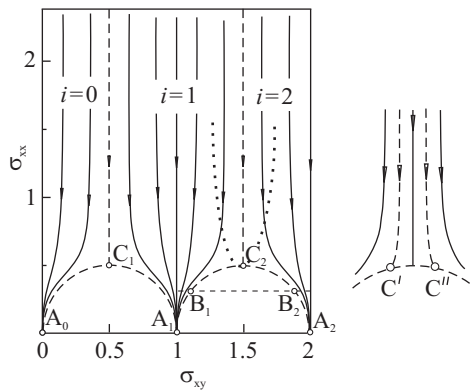}
\caption{Flow diagram for a 2D electron gas in a strong magnetic field
\cite{Khmel1}. The coordinates are represented by the conductivity tensor
components $\sigma_{xy}$ and $\sigma_{xx}$ in the dimensionless units
$e^2/2\pi\hbar$. Separatrices are indicated by dashed lines. A$_i$ are
stationary singular points, C$_i$ are unstable singular points, which are
quantum transition points similar to QCP in Fig. \ref{Qpt5}. The dashed line
indicates the critical region near C$_2$. Horizontal dashed line $B_1$--$B_2$
indicates the image point hopping as the control parameter changes. The
right-hand side of the figure shows the hypothetical flow diagram that
corresponds to a split phase transition and the appearance of a metallic phase}
 \label{Qpt6}
\end{figure}

When eliminating the variable $L$ from two equations (\ref{301}), we find
a relation between $\sigma _{xx}$ and $\sigma _{xy}$, which can be
displayed as curves in the $(\sigma _{xy},\sigma _{xx})$ plane
\cite{Khmel1}. These curves make up a flow diagram for a 2D system of
noninteracting electrons in a strong magnetic field (Fig. \ref{Qpt6}). The
flow lines in this diagram are separated by separatrices periodically
repeated along the $\sigma_{xy}$ axis. This diagram is again a
two-parameter diagram, but due to a strong magnetic field rather than to
interaction. As in the previous flow diagrams, we can move along arrows in
flow trajectories by either increasing the sample size $L$ or decreasing
the temperature at large $L$, i.e., increasing $L_\varphi$.

In the region below points ${\rm C}_i$, the separatrix is split such that
two equivalent points located at the same height $({\rm B}_1$ and ${\rm
B}_2)$ appear in it. The flow lines inside the semicircles in Fig. 17 are
omitted because they are thought to exist separately from the flow lines
outside these semicircles: as the control parameter in the region below
points ${\rm C}_i$ changes, the motion leads to a hop between points
$({\rm B}_1$ and ${\rm B}_2)$ and to a sharp increase in $\sigma _{xy}$.
Strictly speaking, the transfer should occur along a line located outside
the semicircle and bypassing the point ${\rm C}_i$ above. However, for
simplicity, it is indicated by a horizontal dashed line.

The control parameter in the quantum Hall effect regime is usually given
by the electron concentration or the magnetic field. Their effect on the
state of a real system depends on the random field of impurities and other
defects, which transforms discrete Landau levels into minibands and
specifies the energy structure and the character of wave functions in
them. Because a magnetic field determines the magnetic length $r_B=\bigl
(\hbar c/|e|B\bigr)^{1/2}$ as a characteristic scale, we can speak about
two limiting types of random potential, a potential with large-scale
fluctuations of characteristic sizes $\zeta \gg r_B$ and a short-range
potential with $\zeta \ll r_B$. In the long-period potential model, an
energy value $\varepsilon _{\rm c}$ exists near the center of each Landau
miniband such that a delocalized electron wave function corresponds to
this value. If the wave function is strictly delocalized only at
$\varepsilon _{\rm F}=\varepsilon _{\rm c}$ and the random field lifts the
degeneracy of energy levels, a smooth change in the electron concentration
produces a jumplike transformation of the system from one phase state into
another through an isolated energy state with a delocalized wave function
at the Fermi level. This behavior is implied in the flow diagram in Fig.
17, where each isolated metallic state corresponds to a peculiar
separatrix.

The actual width $\delta\varepsilon$ of the energy range with delocalized
wavefunctions depends on finer processes, such as tunneling between two
semiclassical trajectories that are close to each other in the vicinity of
the saddle point (magnetic breakdown). In essence, $\delta\varepsilon$ is
the energy uncertainty of any delocalized state. Another source of
increasing the $\delta\varepsilon$ range is the finiteness of the lengths
$L$ and $L_\varphi$. If the $\delta\varepsilon$ range is finite, a
separatrix is split into two parallel lines and the phase transition is
split into two transitions: a metallic state with a partly filled layer of
extended states at the Fermi level should appear between two Hall-liquid
states whose indices $i$ differ by unity. The corresponding hypothetical
flow diagram is shown on the right-hand side of Fig. 17.

At first glance, it seems that experiment can distinguish between these
two hypothetical possibilities. For definiteness, we assume that the
electron concentration $n$ changes in experiment (considerations for a
change in the magnetic field are similar). As the concentration changes,
the Fermi level moves along the energy scale. When states at the Fermi
level are delocalized, the $\sigma _{xx}$ conductivity is finite, and the
$\sigma _{xy}$ conductivity is in the intermediate region between two
plateaus. Therefore, the temperature dependence of the concentration range
of the intermediate region (i.e., the $\sigma _{xx}$ peak width, the
$\partial \sigma _{xy}/\partial n$ derivative at the center of the
intermediate region, etc.) extrapolated to $T=0$ must determine the energy
range $\delta \varepsilon$ of the delocalized states.

\begin{figure}[h]                       
\includegraphics{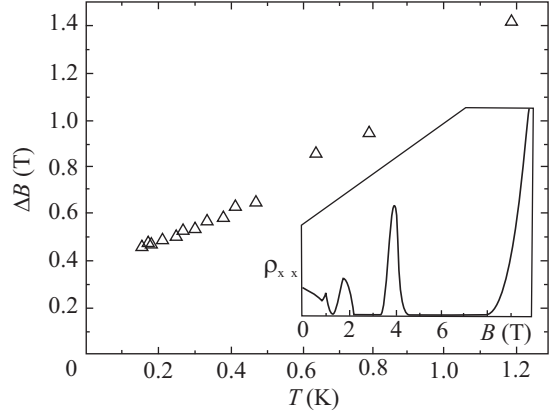}
\caption{Temperature dependence of the peak width of the longitudinal
resistivity $\rho_{xx}$ of a 2D electron gas in the GaAs/AlGaAs heterojunction
during the $2\rightarrow1$ transition in the magnetic field about 4\,�
\cite{Balaban}. The carrier mobility and concentration at $T=1.5$ K are
$\mu=34000$ cm$^2$/V\,s and $n=1.4\cdot10^{11}$ cm$^{-2}$. The inset shows
$\rho_{xx}(B)$ curve recorded at 150 mK }
 \label{Balab}
\end{figure}

However, the experimental results were found to be ambiguous. On the one
hand, the comprehensive experiments in \cite{D1,Balaban} give a finite
value of $\delta\varepsilon$. For example, Fig.\ref{Balab} shows the
transition width measured from the $\rho_{xx}$ resistivity peak width of a
2D gas in the GaAs/AlGaAs heterojunction. The measured function is seen to
be reliably extrapolated to $\Delta B_0\approx0.35$ as $T\rightarrow0$.
However, as we see below, many experiments give the opposite result: as
the temperature decreases, the transition width tends to zero (see, e.g.,
Fig. \ref{Tsui2}). The factor that determines $\delta\varepsilon$ has not
yet been revealed. In any case, there is no unique relation between
$\delta\varepsilon$ and mobility. The statistical characteristics of the
random potential, which are poorly controlled, are likely to play a key
role.

We now move to experiments that do not exhibit a finite energy layer with
delocalized states. Although the flow diagram of $i\rightarrow i+1$
transitions between different states of a Hall liquid has two parameters
(see Fig. \ref{Qpt6}), it is symmetric with respect to both the
$\sigma_{xy}=(i+1/2)(e^{\,2}/2\pi\hbar)$ and
$\sigma_{xy}=i(e^{\,2}/2\pi\hbar)$ axes. Therefore, we may use the version
of the general theory of quantum phase transitions that is based on Eqns
(\ref{qpt16}), (\ref{qpt17}) and assumes that scaling formulas of type
(\ref{qpt21}) are valid. As far as we consider only 2D electronic systems,
all resistances have the same dimension [$\Omega$] and must have the form
\begin{equation}\label{qqeh1}
 R_{uv}=F_{uv}(L_\varphi/\xi)=
 F_{uv}\left(\frac{\delta x}{T^{1/z\nu}}\right),
\end{equation}
in the vicinity of the transition. Here, the $u$ and $v$  subscripts stand
for the $x$ and $y$ coordinates, $F_{uv}$ is an unknown function, and the
argument of the arbitrary $F$ function is written using the last form in
Eqn (\ref{argument}).

\begin{figure}[h]                   
\includegraphics{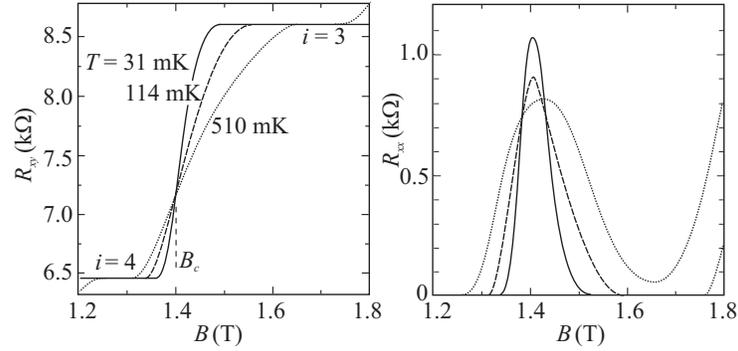}
\caption{Transverse $R_{xy}$ and longitudinal $R_{xx}$ resistances of an
Al$_x$Ga$_{1-x}$As$-$Al$_{0.33}$Ga$_{0.67}$As heterostructure, $x=0.85\%$, at
various temperatures. The critical magnetic field of the 4-3 transition
determined from the point of intersection of the $R_{xy}(T)$ curves is
$B_c=1.40$\,T, \cite{Tsui} } \label{Tsui1}
\end{figure}

\begin{figure}[h]                           
\includegraphics{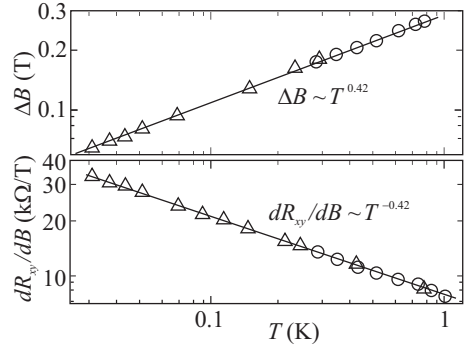}
\caption{Determination of the critical index $\kappa=1/z\nu$ for
transition 4-3 using the $R_{xx}(T)$ peak width (a) and matching the
slopes of the intersecting $R_{xx}(T)$ curves (b). The data were obtained
from  the sample used for Fig. 19 in a dilution refrigerator (triangles)
and a liquid $^3$He cryostat (circles) [33]} \label{Tsui2}
\end{figure}

In contrast to the case of metal--insulator transitions, we analyze
experimental results instead of calculating or predicting the values of
$\nu$ and $z$. As an example, Fig. \ref{Tsui1} shows the magnetic-field
dependence at various temperatures of the longitudinal $R_{xx}$ and
transverse $R_{xy}$ resistances of a Hall bar in a GaAs-based
heterostructure \cite{Tsui}.

According to Eqn (\ref{qqeh1}), the scaling variable
\begin{equation}\label{qqeh5}               
 u=\delta x/T^{1/z\nu}
\end{equation}
is identically zero at all temperatures in the case where the control
parameter takes a critical value; correspondingly, we have
\begin{equation}\label{qqeh6}               
 R_{uv}(x_c,T)={\rm const}.
\end{equation}
Therefore, \emph{separatrix (73) must be horizontal, and all isotherms
$R_{uv}(x,T={\rm const})$ must intersect at one point $x=x_{\rm c}$.} This
is the first test of the applicability of Eqn (\ref{qqeh1}).

We first focus on the $R_{xy}(T\,)$ curves. As is seen from Fig.
\ref{Tsui1}, the $R_{xy}(T\,)$ isotherms obtained for an
Al$_x$Ga$_{1-x}$As--Al$_{0.33}$Ga$_{0.67}$As heterostructure with
$x=0.85\%$ \cite{Tsui} do intersect at one point, $B_{\rm c}=1.40\,$T.
Near the intersection point, all the $R_{xy}(T\,)$ curves can be expanded
into a series and replaced by straight lines $(\partial R_{xy}/\partial
B)_{B_{\rm c}}(B-B_{\rm c})$. As the slopes of these lines are changed
from $(\partial R_{xy}/\partial B)_{B_{\rm c}}$ to $(\partial
R_{xy}/\partial B)_{B_{\rm c}}/T^{\;\kappa}$, where $\kappa =1/z\nu$,
\emph{all the straight lines must merge into one line.} The choice of the
value of $\kappa$ at which the relation
\begin{equation}\label{qqeh7}
 (\partial R_{xy}/\partial B)_{B_c}(T)/T^\kappa={\rm const},
\end{equation}
holds is the second step in the application of the scaling procedure, and
the possibility of this choice is the second condition for the
applicability of the theory. To choose $\kappa$, we plot $(\partial
R_{xy}/\partial B)$ versus $T$ on a $\log \!-\!\log $ scale (Fig.
\ref{Tsui2}).

Formally, Eqn (\ref{qqeh1}) can be applied to both longitudinal ($R_{xx}$)
and transverse ($R_{xy}$) resistances. However, the peak height depends on
the temperature; that is, the point of the maximum does not satisfy
condition (\ref{qqeh6}). Leaving aside the reasons for this fact, we can
use the longitudinal resistance data for scaling analysis by accounting
for the integral property of the $R_{xx}(B)$ functions in the vicinity of
the transition, namely, the peak half-width $\Delta B$ determined by a
certain algorithm. In Fig. \ref{Tsui2}, $\Delta B$ is determined as the
distance between the two maxima of the $(\partial R_{xx}/\partial B)$
derivative.

As can be seen from Fig. \ref{Tsui2}, an analysis of both families of the
functions gives the same value of the critical index $\kappa =0.42$, which
is an additional argument for our scaling procedure.

This value of the critical index $\kappa =0.42$ was repeatedly obtained in
heterostructures made from various materials. However, contrary to
expectations, this value is not universal: other values in the range from
0.2 to 0.8 were also detected in a number of experiments on various
heterostructures. This scatter calls for explanations, since scaling
relations and critical indices are usually universal.

At $T=0$, transition occurs when the condition $\varepsilon _{\rm
F}=\varepsilon _{\rm c}$ is satisfied and therefore the difference
$\delta\varepsilon =|\varepsilon _{\rm F}-\varepsilon _{\rm c}|$ is the
only `internal' control parameter of the system, with the correlation
length $\xi$ depending on it in accordance with a power law:
\begin{equation}\label{qqeh1a}
\xi\propto(\delta\varepsilon)^{\nu_1}.
\end{equation}
From this standpoint, the magnetic field $B$ or the 2D electron
concentration $n$, which depends on the gate voltage $V_{\rm g}$, are
`external' control parameters $x$,
\begin{equation}\label{qqeh2}                   
 \delta x\equiv|B_c-B|\propto(\delta\varepsilon)^{\nu_2}
 \qquad\mbox{or}
 \qquad\delta x\equiv|n_c-n|
 \propto|V_{gc}-V_g|\propto(\delta\varepsilon)^{\nu_2}.
\end{equation}
In both cases, the exponent $\nu _2$ is the same. This is supported by the
fact that the $R_{uv}(B)$ and $R_{uv}(V_{\rm g})$ experimental curves
recorded using the same sample under equivalent conditions differ only in
the scales on the abscissa axis. Eventually, we can write Eqn
(\ref{qpt16}) as
\begin{equation}\label{qqeh3}
 \xi\propto(\delta x)^\nu\qquad(\nu=\nu_1\nu_2),
\end{equation}
where the relation between the control parameter $\delta x$ and the
correlation length $\xi$ consists of the following two links: $\xi$
depends on the position of the Fermi level $\varepsilon _{\rm F}$ with
respect to the delocalized level $\varepsilon _{\rm c}$, and the
difference $\varepsilon _{\rm F}-\varepsilon _{\rm c}$, in turn, depends
on $\delta x$. Correspondingly, according to Eqn (\ref{qqeh3}), the index
$\nu$ turns out to be the product of $\nu _1$ and $\nu _2$. The relation
between $\xi$ and $\delta \varepsilon$  and the related index $\nu _1$ are
likely to be universal and the same for all transitions between different
quantum Hall liquids. But the index $\nu _2$ is determined by the density
of states $g(\varepsilon)$ in the vicinity of the energy $\varepsilon
_{\rm c}$ and, hence, depends on the specific features of the random
potential; this potential can be long- or short-range, statistically
symmetric or asymmetric with respect to the mean value, and so on. The
authors of \cite{Tsui}, whose curves are used in this section, just
studied the effect of specific features of the random potential on
$\kappa$.

The $\Delta B(T\,)$ curve in Fig. \ref{Tsui2} differs radically from the
curve in Fig.\ref{Balab}; in the latter case, a free term $\Delta B_0$ was
introduced for the experimental data to be approximated by a power
function. Nevertheless, we can also perform scaling analysis of the
experimental data in this case using the second hypothetical version of
the flow diagram (see Fig. \ref{Qpt6}). For this interpretation, the
presence of the $\Delta B_0$ term means that in the $2\Delta B_0$ range of
the control parameter $B$, the image point in the flow diagram moves
across the metallic-phase corridor and falls on a separatrix not at $B_0$
corresponding to a maximum of $\sigma _{xx}(B)$ and the derivative
$\partial \sigma _{xy}/\partial B$ but at $B_0+\Delta B_0$. This problem
is discussed in detail in review \cite{Shash}. Here, we only note that
Fig. \ref{Balab} actually contains this scaling analysis. By extrapolating
the $\Delta B(T\,)$ dependence to $T=0$, we can determine the
delocalized-state layer width in units of magnetic field $2\Delta B_0$ and
find that $\Delta B(T\,)-\Delta B_0$ depends linearly on $T$. This means
that $\kappa =1$ in this experiment. The same value of $\kappa$ was
obtained earlier in \cite{D1}.

\section{7. Conclusion}

All the considered cases of metal--insulator transitions were found to be
adequately described by flow diagrams. The list of theoretical works that
have successfully used this technique begins with work \cite{b4}, where a
theoretical model for noninteracting electrons in a zero magnetic field
was developed. The last achievement in this field is the construction of a
flow diagram for a 2D model system of interacting electrons and the
demonstration of the possibility of a metal--insulator transition in this
system .

As regards the general theory of quantum phase transitions, a phase
transition in a 3D system of noninteracting electrons demonstrates that,
in principle, this theory can be used to describe localized--delocalized
electron transitions. Neither conductance, which is used as a physical
quantity specifying the state of the system, nor disorder, which is the
main control parameter, are substantial obstacles for this theory.
However, as usual, various particular cases require theoretical versions
of various degrees of complexity. For example, the version described in
this review cannot be applied to the model proposed in \cite{Fink}.

The relative role and possibilities of both theoretical approaches are
demonstrated when the integer quantum Hall effect is described. The flow
diagram in Fig. \ref{Qpt6} is very convenient for the discussion of the
types of transitions that are possible in a system and for the formulation
of questions to be experimentally checked. Many of these questions are
still open. For example, it is unclear which of the versions of the flow
diagram in Fig. \ref{Qpt6} is realized in reality and whether the
transition between the states with quantum indices $i$ and $i\pm 1$,
\begin{equation}\label{i+1}
i\rightleftarrows i\pm1,
\end{equation}
is split [$i$ is determined by Eqn (\ref{q202})). There are also problems
related to the topology of the flow diagram. According to Fig. \ref{Qpt6},
transitions where quantum number $i$ changes by more than unity are
impossible \cite{Kiv,Huck10}. However, when theorists interpret many
experimental data, they state that such transitions occur (see, e.g.,
review \cite{Shash} and the references therein).

The general theory of quantum phase transitions does not consider the
problem of the relative position of various transitions in the phase
plane. This theory describes the critical vicinity of \emph{one} certain
transition. Proceeding from the assumptions that (a) a transition exists
and (b) the factorization $\xi =\xi(\delta x)$, $L_\varphi =L_\varphi
(T\,)$ occurs in its critical region, the resistivity can be described by
Eqns  (\ref{qpt21})--(\ref{2D-R}) (see Sections 2.3 and 5.3). Then, the
transition point $x=x_{\rm c}$ and critical indices can be determined by
processing the $R(x,T\,)$ curves. This procedure was performed in paper
\cite{Tsui}, which was discussed in Section 6. The value of the results
obtained in \cite{Tsui} becomes clear after the questions formulated above
and following from flow diagrams are answered.

Acknowledgments

We thank A.\ Finkel'shtein and D.\ Khmel'nitskii for the useful
discussions. This work was supported by the Program for the Support of
Leading Scientific Schools (project no. NSh-5930.2006.2) and the Russian
Foundation for Basic Research.

\end{document}